\documentclass[aps,prb,twocolumn,superscriptaddress]{revtex4}
\usepackage{graphics,epsfig,wasysym}

\begin{document}

\title{On the spectrum of fluctuations of a liquid surface:\\
From the molecular scale to the macroscopic scale}

\author{Edgar M. Blokhuis}

\affiliation{Colloid and Interface Science, Leiden Institute of Chemistry,
Gorlaeus Laboratories, P.O. Box 9502, 2300 RA Leiden, The Netherlands.}

\begin{abstract}
We show that to account for the full spectrum of surface fluctuations
from low scattering vector $qd \!\ll\! 1$ (classical capillary wave
theory) to high $qd \!\apprge\! 1$ (bulk-like fluctuations), one must 
take account of the interface's bending rigidity at intermediate
scattering vector $qd \!\apprle\! 1$, where $d$ is the molecular diameter.
A molecular model is presented to describe the bending correction to
the capillary wave model for short-ranged and long-ranged interactions
between molecules. We find that the bending rigidity is negative when the
Gibbs equimolar surface is used to define the location of the fluctuating
interface and that on approach to the critical point it vanishes
proportionally to the interfacial tension. Both features are in agreement
with Monte Carlo simulations of a phase-separated colloid-polymer system.
\end{abstract}

\maketitle

\section{Introduction}

The description of the spectrum of surface fluctuations of a
liquid from the macroscopic scale down to the molecular scale
remains a challenging experimental and theoretical problem.
Using grazing incidence light scattering experiments, Daillant and
coworkers \cite{Daillant} were able, for the first time, to determine
the full spectrum of surface fluctuations, where in previous experiments
(ellipsometry, reflectivity) only certain aspects of the spectrum could
be determined. At the same time, the spectrum can now be analyzed in computer
simulations with ever increasing accuracy \cite{Stecki, Vink, Tarazona}.

Theoretical insight into the structure of a simple liquid surface
is provided by density functional theories on the one hand
\cite{Evans, Henderson, RW} and the capillary wave model on the
other hand \cite{BLS, Weeks, Bedeaux}.
Density functional theories provide a description of the interface
on a microscopic level. The prototype of such theories, the
van der Waals squared-gradient model, was very successful in 
describing, for the first term, the density profile and surface
tension in terms of molecular parameters \cite{RW}. It, however,
fails to capture the subtle role of long wavelength interfacial
fluctuations described by the capillary wave model
\cite{BLS, Weeks, Bedeaux}.

The capillary wave model introduced in 1965 \cite{BLS} describes
the spectrum of fluctuations in terms of a height function
$h(\vec{r}_{\parallel})$ with the surface tension $\sigma$
and gravity $g$ acting as the dominant restoring forces.
The length scale involved in describing capillary waves is the
capillary length, $L_c \equiv \sqrt{\sigma / (m \, \Delta \rho \, g)}$,
which may be as large as a tenth of a millimeter.
The theoretical challenge is to incorporate both theories
and to describe the spectrum of fluctuations of a liquid surface,
as determined from light scattering experiments and computer
simulations, from the molecular scale to the scale of
capillary waves.

An important ingredient in ``bridging the gap'' between
capillary waves and the molecular scale is an extension of the
capillary wave model that incorporates the energy associated
with {\em bending} the interface \cite{Helfrich, Meunier, Blokhuis90}.
Bending is important when
the wavelength of the height fluctuations is approximately
$\sqrt{k_{\rm B} T / \sigma}$, which is typically of the order
of a few times the molecular diameter, i.e. close to the scale
where the molecular structure becomes important and the density
fluctuations are more bulk-like. The natural question that
arises is whether it is possible to describe the full spectrum
of surface fluctuations by the capillary wave model at long
wavelengths and bulk-like fluctuations at the molecular scale.
Is it then necessary to include the leading order correction
to the capillary wave model from bending or are even higher
order terms, relevant at even smaller length scales, required?
 
This article addresses these questions in two parts (a condensed
version has appeared in ref.~\onlinecite{Letter}). In the
first part, we analyze the spectrum of fluctuations recently
obtained by Vink {\em et al.} \cite{Vink} in computer
simulations of a phase-separated polymer-colloid system 
\cite{AOVrij, Gast, Lekkerkerker92, Aarts} in which the
interactions are strictly short-ranged. It is shown that
the simulation data are very accurately described by the combination
of the capillary wave model extended to include a bending
correction, with the bending rigidity as an adjustable parameter,
and bulk-like fluctuations.

In the second part, a molecular basis for the bending correction
to the capillary wave model is offered and the results are
compared with the simulations. The theoretical framework used for
the comparison is mean-field density functional theory in which
the interactions are described by a non-local, integral term
\cite{Evans, Henderson, RW, Mecke}.
The advantage of this approach is that it features the full shape
of the interaction potential enabling the analysis of different
forms and ranges of the interaction potential. We consider
both short-ranged interactions, and long-ranged interactions,
that fall of as $U(r) \!\propto\! 1/r^6$ at large intermolecular
separations.

An important ingredient in our theoretical analysis is the
modification of the density profile, described by $\rho_1(z)$,
due to the local {\em bending} of the interface \cite{Blokhuis93}.
The determination of $\rho_1(z)$ requires one to formulate
precisely the thermodynamic conditions used to vary the
interfacial curvature. Several approaches for the determination
of $\rho_1(z)$ have appeared in the literature
\cite{Mecke, Blokhuis93, Parry94, Blokhuis99}. They differ in
the form of the external field used to set the curvature to a specific
value; in the equilibrium approach \cite{Blokhuis93} the external
field is uniform throughout the system, whereas in the approach
by Parry and Boulter \cite{Parry94, Blokhuis99} it is infinitely
sharp-peaked ($V_{\rm ext} \!\propto\! \delta(z)$) at the interface.
In this article we suggest to add an external field acting
in the interfacial region only with a peak-width of the
order of the thickness of the interfacial region.
The advantage of this approach is that the bulk regions are
unaffected by the additional of the external field and the
resulting $\rho_1(z)$ is a continuous function.

Our paper is organized as follows: in Section 2, the general
form of the surface structure factor to describe the spectrum
of interfacial fluctuations is derived as the combination
of the capillary wave model extended to include a bending
correction and bulk-like fluctuations. This form is then
compared in Section 3 to the Monte Carlo (MC) simulation
results by Vink {\em et al.} \cite{Vink} for the phase-separated
polymer-colloid system. In Section 4, the mean-field density
functional theory used to provide a molecular basis for the
bending extension to the capillary wave model is presented. 
Explicit results are obtained for short-ranged interactions
(Section 5), and long-ranged interactions (Section 6).
We end with a discussion of results.

\section{The fluctuating liquid surface}

In the classical capillary wave model (CW), the fluctuating interface is
described by a two-dimensional surface height function $h(\vec{r}_{\parallel})$,
where $\vec{r}_{\parallel}\!=\!(x,y)$ is the direction parallel to
the surface \cite{BLS, Weeks, Bedeaux}. The fluctuating density
profile can then be written in terms of an ``intrinsic density profile''
shifted over a distance $h(\vec{r}_{\parallel})$:
\begin{equation}
\rho(\vec{r}) = \rho_0(z-h(\vec{r}_{\parallel})) \,,
\end{equation}
where $\rho_0(z)$ is the intrinsic density profile. Often, fluctuations
are assumed to be small so that an expansion in $h$ can be made, neglecting terms
of ${\cal O}(h^2)$,
\begin{equation}
\rho(\vec{r}) = \rho_0(z) - \rho_0^{\prime}(z) \, h(\vec{r}_{\parallel}) + \ldots \,.
\end{equation}
An important consequence of the above linearization is that one may
now identify the intrinsic density profile as the {\em average} density
profile, $\rho_0(z)\!= <\!\rho(\vec{r})\!>$, in view of the fact that
$<\!h(\vec{r}_{\parallel})\!> =\!0$. It is convenient to locate
the $z\!=\!0$ plane such that it coincides with the Gibbs equimolar
surface \cite{RW, Gibbs}, i.e.
\begin{eqnarray}
&& \int\limits_{-\infty}^{\infty} \!\!\! dz \left[ \, <\! \rho(\vec{r}) \!> - \rho_{\rm step}(z) \, \right] \nonumber \\
&& = \int\limits_{-\infty}^{\infty} \!\!\! dz \left[ \, \rho_0(z) - \rho_{\rm step}(z) \, \right] = 0 \,,
\end{eqnarray}
where $\rho_{\rm step}(z)\!=\!\rho_{\ell} \, \Theta(-z) + \rho_v \, \Theta(z)$
with $\Theta(z)$ the Heaviside function and $\rho_{\ell,v}$ the bulk density
in the liquid and vapor region, respectively.

In the above model for $\rho(\vec{r})$, the density correlations are
essentially given by the correlations of $h(\vec{r}_{\parallel})$,
which are described by the height-height correlation function:
\begin{equation}
S_{hh}(r_{\parallel}) \equiv \,\, <\!h(\vec{r}_{1,\parallel}) \, h(\vec{r}_{2,\parallel})\!> \,,
\end{equation}
where $\vec{r}_{\parallel}\!\equiv\!\vec{r}_{2,\parallel} - \vec{r}_{1,\parallel}$
and $r_{\parallel}\!\equiv \mid\!\vec{r}_{\parallel}\!\mid$.

To determine the height-height correlation function, one should
examine the change in free energy, $\Delta \Omega$, associated with a
fluctuation of the interface. In the capillary wave model it is
described by considering the change in free energy associated with a
distortion of the surface against gravity and surface area extension
\cite{BLS}:
\begin{equation}
\Delta \Omega = \frac{1}{2} \int \!\! d\vec{r}_{\parallel}
\left[ m \, \Delta \rho \, g \, h(\vec{r}_{\parallel})^2 +
\sigma \, | \vec{\nabla} h(\vec{r}_{\parallel}) |^2 \right] \,.
\end{equation}
It is convenient to express $\Delta \Omega$ in terms of the
Fourier Transform of $h(\vec{r}_{\parallel})$, $h(\vec{q})\!=\!\int \! d\vec{r}_{\parallel} \,
e^{- i \vec{q} \cdot \vec{r}_{\parallel}} \, h(\vec{r}_{\parallel})$,
\begin{equation}
\label{eq:Omega_CW}
\Delta \Omega = \frac{1}{2} \int \!\! \frac{d\vec{q}}{(2 \pi)^2} \;
\left[m \, \Delta \rho \, g + \sigma \, q^2 \right] \, h(\vec{q}) \, h(-\vec{q}) \,.
\end{equation}
In the capillary wave model the height-height correlation function is determined
by a full Statistical Mechanical analysis \cite{Weeks, Bedeaux} in which the above
expression for the change in free energy is interpreted as the so-called capillary
wave Hamiltonian, $\Delta \Omega\!=\!{\cal H}_{cw}[h(\vec{r}_{\parallel})]$.
In general, one has
\begin{equation}
S_{hh}(r_{\parallel}) = \frac{1}{Z} \int \!\! {\cal D}h \;
h(\vec{r}_{1,\parallel}) \, h(\vec{r}_{2,\parallel}) \,
e^{-{\cal H}_{cw}[h]/k_{\rm B} T} \,,
\end{equation}
where $Z$ is the partition function associated with ${\cal H}_{cw}[h]$,
$k_{\rm B}$ is Boltzmann's constant and $T$ is the temperature.
It can be shown that \cite{Weeks, Bedeaux}
\begin{eqnarray}
S_{hh}(q) &=& \int \!\! d\vec{r}_{\parallel} \,
e^{- i \vec{q} \cdot \vec{r}_{\parallel}} \, S_{hh}(r_{\parallel}) \\
&=& \frac{k_{\rm B} T}{m \, \Delta \rho \, g + \sigma \, q^2}
= \frac{k_{\rm B} T}{\sigma \, (L_c^{-2} + q^2)} \,. \nonumber
\end{eqnarray}
For simplicity, we ignore gravity effects
in the following and set $L_c\!=\!\infty$ ($g\!=\!0$).

\subsection{Extended capillary wave model}

In the derivation of the classical capillary wave model, one
assumes an expansion in gradients of $h(\vec{r}_{\parallel})$,
$\mid\! \vec{\nabla} h \!\mid \ll 1$.
In the {\em extended} capillary wave model (ECW), one wishes to extend
the expansion by including higher derivatives of $h(\vec{r}_{\parallel})$.
To leading order one may then write the fluctuating density as
\cite{Blokhuis99, Mecke}
\begin{equation}
\label{eq:rho_ECW}
\rho(\vec{r}) = \rho_0(z) - \rho^{\prime}_0(z) \, h(\vec{r}_{\parallel})
- \frac{\rho_1(z)}{2} \, \Delta h(\vec{r}_{\parallel}) + \ldots
\end{equation}
The function $\rho_1(z)$ is identified as the correction to the
density profile due to the {\em curvature} of the interface,
$\Delta h(\vec{r}_{\parallel})\!\approx\!-1/R_1-1/R_2$, with $R_1$
and $R_2$ the (principal) radii of curvature. The prefactor of $-1/2$
is chosen such that the notation is consistent with an analysis in
which the curvature does not result from a fluctuation of the planar
interface, but is due to the fact that one considers a spherical
liquid droplet ($R_1\!=\!R_2\!=\!R$) in (metastable) equilibrium
with a bulk vapor phase \cite{Blokhuis92, Blokhuis93, Gompper}.
An expansion in the curvature of the density profile $\rho_s(r)$
then gives
\begin{equation}
\rho_s(r) = \rho_0(r) + \frac{\rho_1(r)}{R} + \ldots
\end{equation}
which parallels the expansion in Eq.(\ref{eq:rho_ECW}).

The inclusion of curvature corrections in the extended capillary wave model
leads to higher order terms in an expansion in $q^2$, terms beyond
$\sigma q^2$, in the expression for $\Delta \Omega$ in Eq.(\ref{eq:Omega_CW}).
It is customary to capture these higher order terms by introducing a
wave vector dependent surface tension $\sigma(q)$ \cite{Meunier}
\begin{equation}
\label{eq:Omega_ECW}
\Delta \Omega = \frac{1}{2} \int \!\! \frac{d\vec{q}}{(2 \pi)^2} \;
\sigma(q) \, q^2 \, h(\vec{q}) \, h(-\vec{q}) \,,
\end{equation}
which gives for the height-height correlation function
\begin{equation}
\label{eq:S_hh_ECW}
S_{hh}(q) = \frac{k_{\rm B} T}{\sigma(q) \, q^2} \,.
\end{equation}
The precise form of $\sigma(q)$ depends sensitively on the behavior
of the interaction potential at large distances \cite{Mecke}. When
the interaction potential is sufficiently short-ranged (SR), the
expansion of $\sigma(q)$ in $q^2$ is regular and the leading correction
is of the form:
\begin{equation}
\label{eq:sigma_ECW_SR}
\sigma(q) = \sigma + k \, q^2 + {\cal O}(q^4) \,. \hspace*{25pt} {\rm (SR)}
\end{equation}
The coefficient $k$ is identified as the {\em bending rigidity}
\cite{Helfrich, Meunier, Blokhuis90}.
This is because the form for $\Delta \Omega$ in Eq.(\ref{eq:Omega_ECW}),
with $\sigma(q)$ given by Eq.(\ref{eq:sigma_ECW_SR}), can also be
derived from the Helfrich free energy expression \cite{Helfrich},
which reads for a fluctuating interface:
\begin{equation}
\Delta \Omega = \frac{1}{2} \int \!\! d\vec{r}_{\parallel}
\left[ \sigma \, | \vec{\nabla} h(\vec{r}_{\parallel}) |^2
+ k \left( \Delta h(\vec{r}_{\parallel}) \right)^2 \right] \,.
\end{equation}

When the interaction potential is long-ranged (LR), specifically when
it falls of as $U(r)\!\propto\!1/r^6$ at large intermolecular distances,
which is the case for regular fluids due to London-dispersion forces,
one finds that the leading correction to $\sigma(q)$ picks up a
logarithmic contribution \cite{Mecke}:
\begin{equation}
\label{eq:sigma_ECW_LR}
\sigma(q) = \sigma + k_s \, q^2 \, \ln(q \ell_k) + {\cal O}(q^4) \,,
\hspace*{8pt} {\rm (LR)}
\end{equation}
with $k_s$ and $\ell_k$ parameters independent of $q$. The coefficient
$k_s$ depends on the asymptotic behavior of $U(r)$ but is otherwise
a {\em universal} constant \cite{Mecke}. The bending length $\ell_k$ depends,
like the bending rigidity $k$, on the microscopic parameters of the model.
In principal, all the parameters $\sigma$, $k$, $k_s$, and $\ell_k$
can be expressed in terms of the density profiles $\rho_0(z)$ and
$\rho_1(z)$ by inserting the fluctuating density as given in
Eq.(\ref{eq:rho_ECW}) into a microscopic model for the free energy
and comparing the result with Eq.(\ref{eq:Omega_ECW}).

It is important to realize that the extended capillary wave model
assumes a curvature expansion in Eq.(\ref{eq:rho_ECW}) which
translates into an expansion in $q^2$ in Eq.(\ref{eq:Omega_ECW}) that
is valid only up to ${\cal O}(q^4)$. Higher order terms are not
systematically included. The result is that one should limit the
expansion of $\sigma(q)$ in Eq.(\ref{eq:sigma_ECW_SR}) or
Eq.(\ref{eq:sigma_ECW_LR}) to the order in $q$ indicated.

\subsection{Definition of the height profile}

An important subtlety in the preceding analysis is the fact that the
location of the interface, i.e. the value of the height function
$h(\vec{r}_{\parallel})$, cannot be defined {\em unambiguously} \cite{Gibbs}.
A certain procedure must always be formulated to determine $h(\vec{r}_{\parallel})$.
It turns out that the choice for $h(\vec{r}_{\parallel})$
influences the density profile $\rho_1(z)$ which, in turn,
determines the value of the bending parameters $k$ and $\ell_k$.

We explicitly consider two canonical choices for the determination
of $h(\vec{r}_{\parallel})$; the {\em crossing} constraint (cc)
and the {\em integral} constraint (ic) \cite{Parry94, Blokhuis99}.
Other choices are certainly possible and equally legitimate as
long as they lead to a location of the dividing surface that is
`sensibly coincident' with the interfacial region \cite{Gibbs}.
In this context we like to mention the work by Tarazona {\em et al.}
\cite{Tarazona}, who propose a `state of the art' manner to define
the location of the interface based on the distribution of molecules
rather than the molecular density alone. 

In the {\em crossing constraint}, $h(\vec{r}_{\parallel})$ is defined
as the height where the fluctuating density equals some fixed value
of the density that lies in between the limiting bulk densities,
say $\rho(\vec{r}) \!=\! \rho_0(z\!=\!0)$:
\begin{equation}
\rho(\vec{r}_{\parallel},z\!=\!h(\vec{r}_{\parallel})) = \rho_0(0) \,.
\hspace*{20pt} {\rm (cc)}
\end{equation}
Using this condition in Eq.(\ref{eq:rho_ECW}), one finds the following
constraint for $\rho_1(z)\!=\!\rho^{cc}_1(z)$
\begin{equation}
\rho^{cc}_1(0) = 0 \,.
\end{equation}
In the {\em integral constraint}, $h(\vec{r}_{\parallel})$ is defined
by the integral over the fluctuating density \cite{Gibbs}
\begin{equation}
h(\vec{r}_{\parallel}) = \frac{1}{\Delta \rho} \int\limits_{-\infty}^{\infty} \!\!\! dz
\left[ \, \rho(\vec{r}) - \rho_{\rm step}(z) \, \right] \,.
\hspace*{8pt} {\rm (ic)}
\end{equation}
With this condition inserted into Eq.(\ref{eq:rho_ECW}), one now finds
that $\rho_1(z)\!=\!\rho^{ic}_1(z)$ is subject to the following constraint
\begin{equation}
\int\limits_{-\infty}^{\infty} \!\!\! dz \, \rho^{ic}_1(z) = 0 \,.
\end{equation}
We show in Section 4 that the ambiguity in locating the
dividing surface translates into the density profile $\rho_1(z)$ being
determined up to an additive factor proportional to $\rho_0^{\prime}(z)$ 
\cite{Blokhuis99}. In particular, $\rho^{ic}_1(z)$ and $\rho^{cc}_1(z)$
are related by
\begin{equation}
\label{eq:relation_rho1_ic_cc}
\rho^{ic}_1(z) = \rho^{cc}_1(z) + \alpha \, \rho_0^{\prime}(z) \,.
\end{equation}
The value of the constant $\alpha$ can be determined by integrating
both sides of the above equation over $z$
\begin{equation}
\label{eq:alpha}
\alpha = \frac{1}{\Delta \rho} \int\limits_{-\infty}^{\infty} \!\!\! dz \, \rho^{cc}_1(z) \,.
\end{equation}
One may further show that the ambiguity in the determination of $\rho_1(z)$
is of influence to the value of the bending parameters $k$ and $\ell_k$.
In Section 4 we show that because $\rho^{ic}_1(z)$ and $\rho^{cc}_1(z)$ are
related by Eq.(\ref{eq:relation_rho1_ic_cc}), we have for the bending parameters
\cite{Blokhuis99}
\begin{eqnarray}
\label{eq:k_shift}
k^{ic} &=& k^{cc} - \alpha \, \sigma \,, \nonumber \\
k_s \, \ln(\ell_k^{ic}) &=& k_s \, \ln(\ell_k^{cc}) - \alpha \, \sigma \,.
\end{eqnarray}
Naturally, all experimentally measurable quantities {\em cannot} depend on
the choice made for the location of the height function $h(\vec{r}_{\parallel})$.
The implication is that it is necessary to formulate precisely
the quantity that is determined experimentally and verify that its
value is independent of the choice for $h(\vec{r}_{\parallel})$.
This is explicitly shown next.

The quantity studied in experiments and simulations is the (surface) density-density
correlation function. It is an integral into the bulk region to a certain depth
$L$ of the density-density correlation function:
\begin{eqnarray}
\label{eq:definition_S(r)}
&& S(r_{\parallel}) \equiv \frac{1}{(\Delta \rho)^2} \int\limits_{-L}^{L} \!\! dz_1
\int\limits_{-L}^{L} \!\! dz_2 \\
&& \hspace*{5pt} <\! [ \rho(\vec{r}_1) - \rho_{\rm step}(z_1) ] \,
[ \rho(\vec{r}_2) - \rho_{\rm step}(z_2) ] \!> \,. \nonumber
\end{eqnarray}
When we insert the general expression for $\rho(\vec{r})$ as given by
Eq.(\ref{eq:rho_ECW}) into Eq.(\ref{eq:definition_S(r)}), one finds that
\begin{eqnarray}
\label{eq:S(r)}
S(r_{\parallel}) &=& \,\, <\! h(\vec{r}_{1,\parallel}) \, h(\vec{r}_{2,\parallel}) \!> \\
&& - \, \frac{1}{\Delta \rho} \int\limits_{-\infty}^{\infty} \!\!\! dz \, \rho_1(z)
<\! h(\vec{r}_{1,\parallel}) \, \Delta h(\vec{r}_{2,\parallel}) \!> \nonumber
\end{eqnarray}
where we can neglect a term $<\! \Delta h \, \Delta h \!>$ to the
order in the curvature expansion considered. Furthermore, we have
assumed that $L$ is sufficiently large so that we can approximate
\begin{eqnarray}
\label{eq:L_large}
\int\limits_{-L}^{L} \!\! dz \, \rho^{\prime}_0(z) &\approx& - \Delta \rho \,, \nonumber \\
\int\limits_{-L}^{L} \!\! dz \, \rho_1(z) &\approx& 
\int\limits_{-\infty}^{\infty} \!\!\! dz \, \rho_1(z) \,.
\end{eqnarray}
Rather than $S(r_{\parallel})$, we consider its Fourier Transform,
$S(q)$, which we shall term the {\em surface structure factor}:
\begin{eqnarray}
\label{eq:S(q)}
S(q) &=& \int \!\! d\vec{r}_{\parallel} \,
e^{- i \vec{q} \cdot \vec{r}_{\parallel}} \, S(r_{\parallel}) \\
&=& S_{hh}(q) + \, \frac{1}{\Delta \rho}
\int\limits_{-\infty}^{\infty} \!\!\! dz \, \rho_1(z) \, q^2 \, S_{hh}(q) \,. \nonumber
\end{eqnarray}
We now verify that $S(q)$ is {\em independent} of the choice for
$h(\vec{r}_{\parallel})$ by determining $S(q)$ using both the
integral constraint and crossing constraint. For simplicity,
we consider the case of short-ranged forces only (the verification
for the case of long-ranged forces follows analogously).
The surface structure factor using both constraints is given by
\begin{eqnarray}
S^{cc}(q) &=& \frac{k_{\rm B} T}{\sigma \, q^2 + k^{cc} \, q^4 + ..}
+ \frac{\alpha \, k_{\rm B} T \, q^2}{\sigma \, q^2 + k^{cc} \, q^4 + ..} \nonumber \\
&=& \frac{k_{\rm B} T}{\sigma \, q^2} - \frac{k_{\rm B} T \, k^{cc}}{\sigma^2} 
+ \frac{\alpha \, k_{\rm B} T}{\sigma} + {\cal O}(q^2) \,, \nonumber \\
S^{ic}(q) &=& \frac{k_{\rm B} T}{\sigma \, q^2 + k^{ic} \, q^4 + ..} \nonumber \\
&=& \frac{k_{\rm B} T}{\sigma \, q^2} - \frac{k_{\rm B} T \, k^{ic}}{\sigma^2} + {\cal O}(q^2) \,,
\end{eqnarray}
where we have used the explicit expression for $S_{hh}(q)$ in
Eq.(\ref{eq:S_hh_ECW}) together with Eq.(\ref{eq:sigma_ECW_SR}). On account
of the fact that $k^{ic}\!=\!k^{cc} - \alpha \sigma$, one finds that
$S^{cc}(q) \!=\! S^{ic}(q) \!\equiv\! S(q)$ as required.

This analysis shows that $S(q)$ equals the height-height
correlation function when the {\em integral} constraint is used
to define the location of the height profile, i.e.
\begin{equation}
S(q) = S^{ic}_{hh}(q) \,.
\end{equation}
It is therefore convenient, but by no means necessary, to use
the integral constraint to define the location of the dividing surface.

Finally, we consider the contribution of ``bulk-like'' fluctuations
to the fluctuating density profile which are predominantly present
at short wavelengths, $qd \!\apprge\! 1$. 

\subsection{Bulk-like fluctuations}

\begin{figure}
\centering
\includegraphics[angle=270,width=200pt]{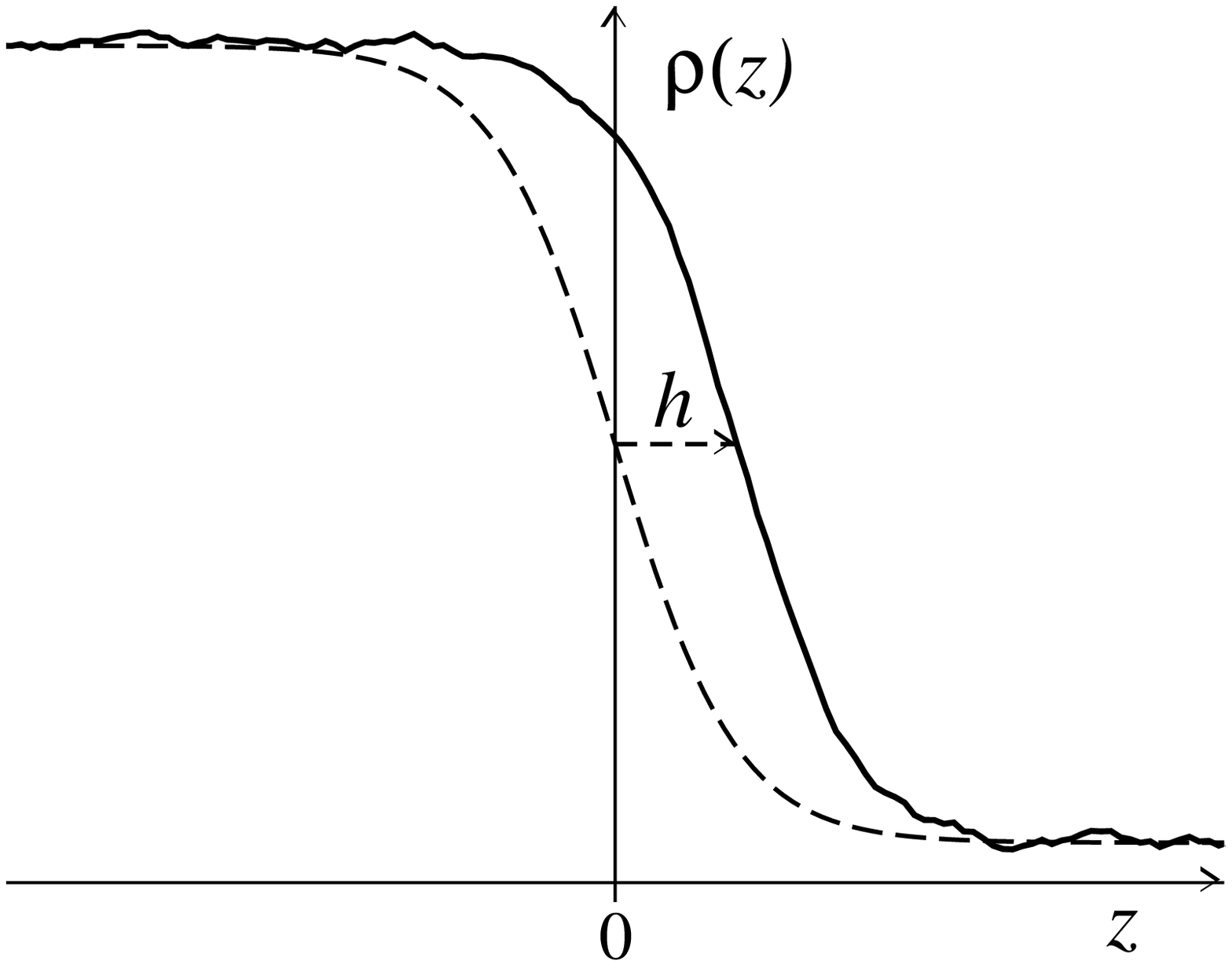}
\caption{Sketch of the fluctuating density profile as a function
of $z$; the height $h\!=\!h(\vec{r}_{\parallel})$ is the distance
over which the intrinsic density profile $\rho_0(z)$ (dashed line)
is shifted.}
\label{fig:fluct_rho}
\end{figure}
Adding short wavelength, bulk-like fluctuations to the fluctuating density,
the full picture that emerges for $\rho(\vec{r})$ is that schematically
depicted in Figure \ref{fig:fluct_rho}. It can be described as:
\begin{equation}
\label{eq:rho_ECWB}
\rho(\vec{r}) = \rho_0(z) - \rho^{\prime}_0(z) \, h(\vec{r}_{\parallel})
- \frac{\rho_1(z)}{2} \, \Delta h(\vec{r}_{\parallel}) + \delta \rho_b(\vec{r})
\end{equation}
where $\delta \rho_b(\vec{r})$ represents the bulk-like fluctuations.
We shall consider only small fluctuations so that $<\!\delta \rho_b\!>=\!0$
and assume that there are no correlations between height fluctuations
and bulk-like fluctuations, $<\! h \, \delta \rho_b \!> = \! 0$.
When we insert the expression for $\rho(\vec{r})$ as given by Eq.(\ref{eq:rho_ECWB})
into the expression for $S(r_{\parallel})$ in Eq.(\ref{eq:definition_S(r)}),
one finds that
\begin{eqnarray}
\label{eq:S(r)_B}
&& S(r_{\parallel}) = \,\, S^{ic}_{hh}(r_{\parallel}) \\
&& \hspace*{5pt} + \frac{1}{(\Delta \rho)^2} \, \int\limits_{-L}^{L} \!\! dz_1
\! \int\limits_{-\infty}^{\infty} \!\!\! dz_{12}
<\! \delta \rho_b(\vec{r}_1) \, \delta \rho_b(\vec{r}_2) \!> \,. \nonumber
\end{eqnarray}
Here we have made a further approximation by replacing the integration over
$z_2$ from $-L$ to $L$ by an integral over $z_{12}$ from $-\infty$ to $\infty$.
The integral over $z_1$ that is left gives rise to a term that increases
linearly with $L$. That means that the bulk-like contributions to
$S(r_{\parallel})$ eventually dominate the height fluctuations when
$L$ becomes larger. To study surface fluctuations via $S(r_{\parallel})$
it is therefore important that on the one hand $L$ is sufficiently large
in order to make the approximations in, e.g., Eq.(\ref{eq:L_large})
but on the other hand not so large as to completely dominate the
contribution from surface height fluctuations.
In the next section we show how these two conditions pan out for the
circumstances under which the simulation results are obtained.

A further issue is that the bulk density correlation function
$<\! \delta \rho_b \, \delta \rho_b \!>$ differs in either phase
(liquid or vapor). When one then considers the integral over $z_1$,
it seems appropriate to approximate $<\! \delta \rho_b \, \delta \rho_b \!>$
by the density correlation function in the {\em bulk liquid} region:
\begin{equation}
<\! \delta \rho_b(\vec{r}_1) \, \delta \rho_b(\vec{r}_2) \!> \, = \rho_{\ell}^2 \,
\left[ \, g_{\ell}(r) - 1 \, \right] + \rho_{\ell} \, \delta(\vec{r}_{12}) \,,
\end{equation}
and introduce an $L$-dependent prefactor ${\cal N}_L$ to account for
the integral over $z_1$. The surface structure factor thus becomes
\begin{equation}
\label{eq:S(q)_B}
S(q) = S^{ic}_{hh}(q) + {\cal N}_L \, S_b(q) \,,
\end{equation}
with the bulk structure factor $S_b(q)$ defined as
\begin{equation}
S_b(q) = 1 + \rho_{\ell} \int \!\! d\vec{r}_{12} \,
e^{- i \vec{q} \cdot \vec{r}_{12}} \; \left[ \, g_{\ell}(r) - 1 \, \right] \,.
\end{equation}
This approximation may be justified by arguing that close to the
critical point there is no distinction between the two bulk correlation
functions, whereas far from the critical point the contribution
from the bulk vapor can be neglected since $\rho_v\!\approx\!0$.

The value for the $L$-dependent prefactor ${\cal N}_L$ may be determined
from a fit to the limiting behavior of $S(q)$ at $q \rightarrow\! \infty$.
For an explicit evaluation of $S_b(q)$, we have taken for $g_{\ell}(r)$ the
Percus-Yevick solution \cite{PY} for the hard-sphere correlation
function, $g_{\ell}(r)\!=\!g_{\rm hs}^{\rm PY}(r;\rho_{\ell})$. 

\section{Comparison with Monte Carlo simulations}

In this section, the surface structure factor in Eq.(\ref{eq:S(q)_B})
is compared to results from Monte Carlo simulations by Vink {\em et al.}
\cite{Vink}. The system considered consists of a mixture
of colloidal particles with diameter $d$ and polymer particles with
diameter $2 R_{\rm g}$. The colloid-colloid and colloid-polymer
interactions are considered to be hard-sphere like, whereas polymer-polymer
interactions are taken ideal. The presence of polymer induces a depletion
attraction between the colloidal particles which may ultimately lead to
phase separation \cite{AOVrij, Gast, Aarts, Lekkerkerker92}. The resulting interface
of the demixed colloid-polymer system is studied by Vink {\em et al.}
\cite{Vink} for a number of polymer concentrations and for a polymer-colloid
size ratio parameter $\varepsilon \!\equiv\! 1 + 2 R_{\rm g}/d\!=\!$ 1.8.

To study the interfacial fluctuations, Vink {\em et al.} introduce the
local interface position as \cite{Vink}:
\begin{equation}
z_G(\vec{r}_{\parallel}) \equiv \frac{1}{\Delta \rho} \int\limits_{-L}^{L} \!\! dz
\left[ \, \rho(\vec{r}) - \rho_{\rm step}(z) \, \right] \,,
\end{equation}
where $\rho(\vec{r})$ can be taken to be either the colloid or polymer density.
The integration limits $\pm L$ are inside the bulk regions, but different values
for it are systematically considered \cite{Vink}. One may easily verify that
the correlations of the local interface position are exactly described by the
surface structure factor defined earlier in Eq.(\ref{eq:definition_S(r)})
\begin{equation}
<\!z_G(\vec{r}_{1,\parallel}) \, z_G(\vec{r}_{2,\parallel}) \!> \, = \, S(r_{\parallel}) \,.
\end{equation}
\begin{figure}
\centering
\includegraphics[angle=270,width=200pt]{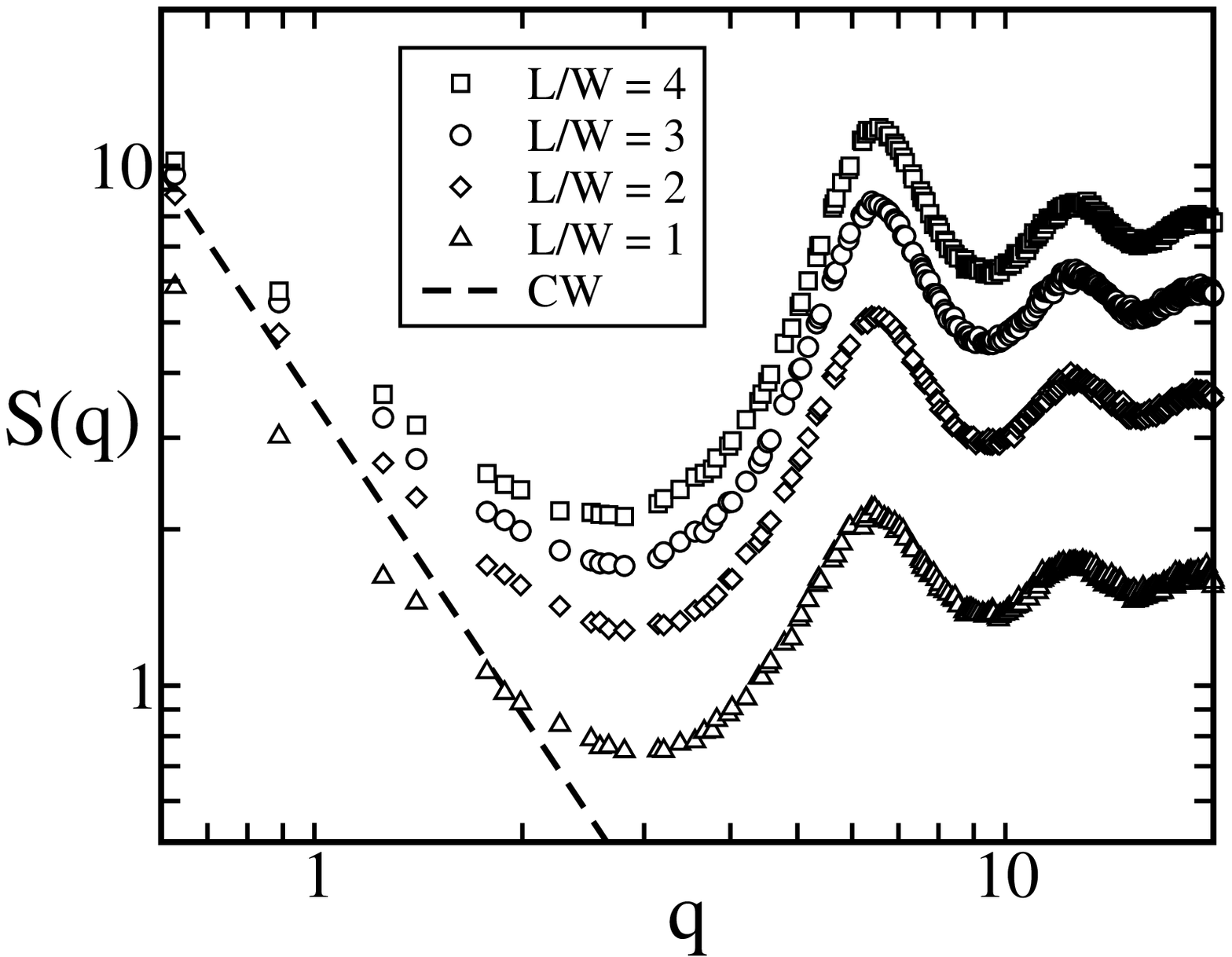}
\caption{MC results by Vink (Ref.~\onlinecite{Vink}) for the surface structure factor
(in units of $d^4$) versus $q$ (in units of $1/d$) for various values
of the integration limit $L/W\!=\!$ 1, 2, 3, 4. The dashed line is the
capillary wave model. In this example $\varepsilon\!=\!$ 1.8,
$\eta_{p}\!=\!$ 1.0, and the colloidal particles are used to define $z_G$.}
\label{fig:Vink0}
\end{figure}
\noindent
In Figure \ref{fig:Vink0}, typical results for the Fourier transform of the
surface structure obtained in the MC simulations of Vink
are shown (Figure 13 of ref.~\onlinecite{Vink}). In this example the integration limit
is varied, $L/W\!=\!$ 1, 2, 3, 4, where $W$ is some measure of the interfacial
thickness. One clearly observes that when $L/W$ is too small, the results
do not match the classical capillary wave behavior for small $q$ (dashed line),
and that the contribution from bulk-like fluctuations at high $q$ increases
with $L/W$.

\begin{figure}
\centering
\includegraphics[angle=270,width=200pt]{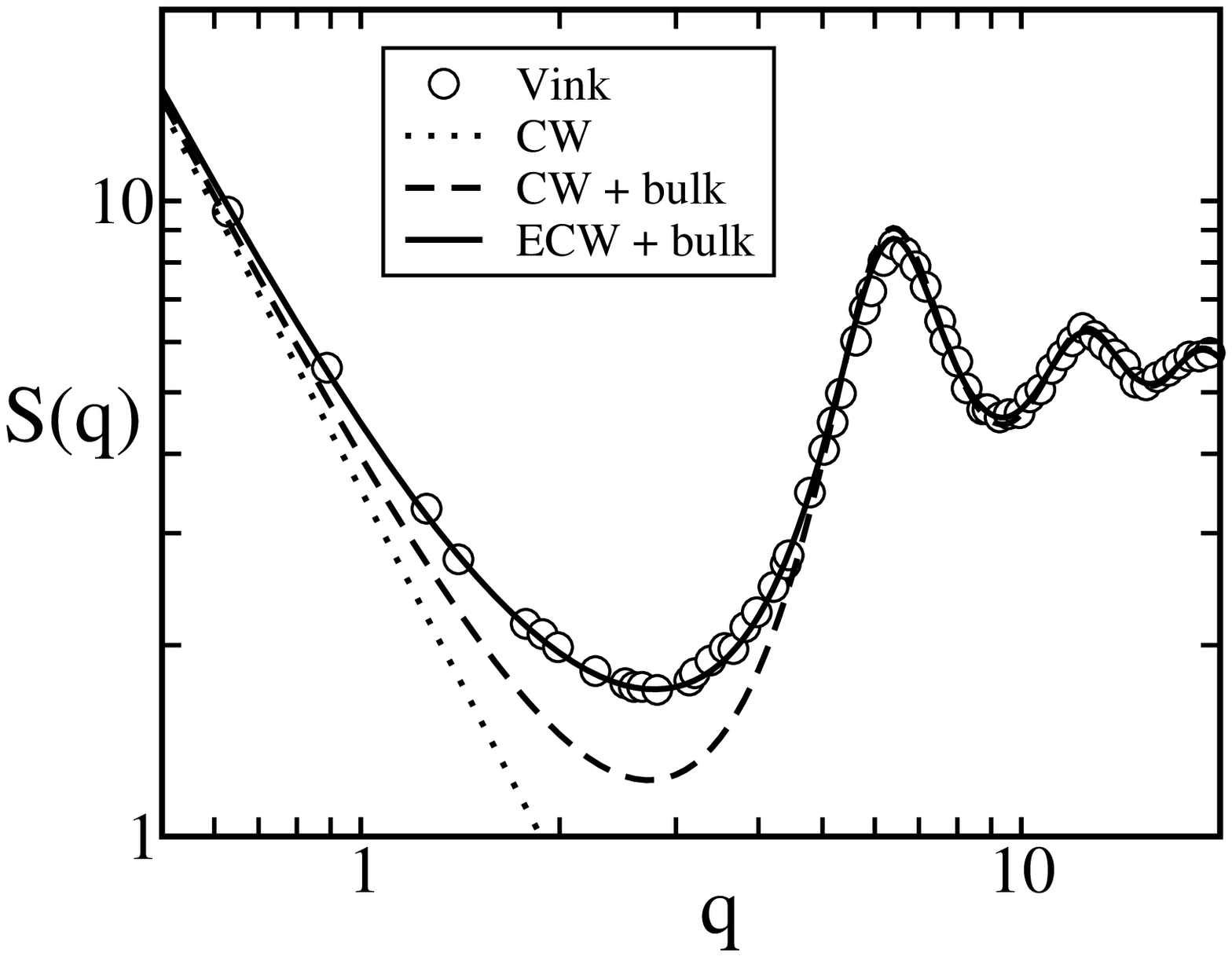}
\caption{MC results by Vink {\em et al.} (Ref.~\onlinecite{Vink}) for the surface structure factor
(in units of $d^4$) versus $q$ (in units of $1/d$). The dotted line is the
capillary wave model, the dashed line is the combination of the capillary wave
model and the bulk correlation function, and the drawn line is the combination
of the extended capillary wave model and the bulk correlation function. In this
example $\varepsilon\!=\!$ 1.8, $\eta_{p}\!=\!$ 1.0, $L/W\!=\!$ 3, and the
colloidal particles are used to define $z_G$.}
\label{fig:Vink1}
\end{figure}

In Figure \ref{fig:Vink1}, we consider the result from Figure \ref{fig:Vink0}
for $L/W\!=\!$ 3. For small $q$ the results asymptotically approach the
result of the classical capillary wave model (dotted line) with the value
of $\sigma$ taken from separate simulations. The dashed line is the
combination of the capillary wave model with the bulk correlation function:
\begin{equation}
\label{eq:S(q)_CWB}
S(q) = \frac{k_{\rm B} T}{\sigma \, q^2} + {\cal N}_L \, S_b(q) \,. 
\end{equation}
The value of ${\cal N}_L$ is chosen such that it matches the $q\!\rightarrow\!\infty$
limit for $S(q)$ in Figure \ref{fig:Vink1}. One finds that Eq.(\ref{eq:S(q)_CWB})
already matches the simulation results quite accurately except at intermediate
values of $q$, $qd \!\approx\! 1$.

As a next step, we investigate whether the inclusion of a bending rigidity
is able to describe the simulation results at these intermediate values:
\begin{equation}
\label{eq:S(q)_ECWB1}
S(q) = \frac{k_{\rm B} T}{\sigma \, q^2 + k^{ic} \, q^4 + ..}
+ {\cal N}_L \, S_b(q) \,. 
\end{equation}
The bending rigidity describes the leading order correction to the
classical capillary wave model in an expansion in $q^2$. Its value
is therefore obtained from analyzing the behavior of $S(q)$ when
$qd \!\apprle\! 1$. The fact that the simulation results in Figure \ref{fig:Vink1}
are systematically {\em above} the capillary wave prediction in
this region, indicates that the bending rigidity thus obtained is {\em negative},
$k^{ic} \!<\! 0$. Unfortunately, a negative bending rigidity prohibits the
use of Eq.(\ref{eq:S(q)_ECWB1}) to fit the simulation results in the
{\em entire} $q$-range since the denominator becomes zero at a certain
value of $q$. It is therefore convenient to rewrite the expansion in $q^2$
in Eq.(\ref{eq:S(q)_ECWB1}) in the following form:
\begin{equation}
\label{eq:S(q)_ECWB}
S(q) = \frac{k_{\rm B} T}{\sigma \, q^2} \,
(1 - \frac{k^{ic}}{\sigma} \, q^2 + \ldots) + {\cal N}_L \, S_b(q) \,,
\end{equation}
which is equivalent to Eq.(\ref{eq:S(q)_ECWB1}) to the order in $q$ considered,
but which has the advantage of being well-behaved in the entire $q$-range.
Other forms to regulate $S(q)$, that are equivalent to Eq.(\ref{eq:S(q)_ECWB1})
to the order in $q$ considered, may certainly be formulated.
In analogy with a similar treatment of capillary waves by Parry
and coworkers \cite{Parry08} in the context of wetting transitions, one might
suggest that the appearance of a negative bending rigidity indicates
the missing of a correlation length that would replace Eq.(\ref{eq:S(q)_ECWB})
with an explicit formula valid for all values of $q$, not just
to the order in $q$ considered.

The above form for $S(q)$ in Eq.(\ref{eq:S(q)_ECWB}), with the bending rigidity
used as an adjustable parameter ($k \!=$ - 0.045 $k_{\rm B} T$), is plotted
in Figure \ref{fig:Vink1} as the drawn line. Exceptionally good agreement with
the Monte Carlo simulations is obtained. In Table 1, we list values of the bending
rigidity obtained for a number of polymer volume fractions, $\eta_p$. These values
are the results of fits of $S(q)$ from Monte Carlo simulations for several system sizes
and for several values of $L/W$, with the error estimated from the standard deviation
of the various results. For the $L/W \!=\!1$ and $L/W \!=\!2$ curves
(see Figure \ref{fig:Vink0}), one needs to adjust for the fact that the capillary wave
limit is not correctly approached at low $q$. For the results in Figure \ref{fig:Vink0},
one ultimately obtains for the bending rigidity $k \!=$ - 0.040, - 0.040, - 0.045,
- 0.060 $k_{\rm B} T$, for $L/W \!=\!$ 1, 2, 3, and 4, respectively.

In Table 1, it should be reminded that, rather than the true polymer volume fraction
in either phase, $\eta_p$ should be interpreted as the polymer volume fraction of a
reservoir fixing the polymer chemical potential \cite{Lekkerkerker90}. Furthermore,
the ``liquid'' is defined as the phase relatively rich in colloids and the ``vapor''
as the phase relatively poor in colloids.

\begin{figure}
\centering
\includegraphics[angle=270,width=200pt]{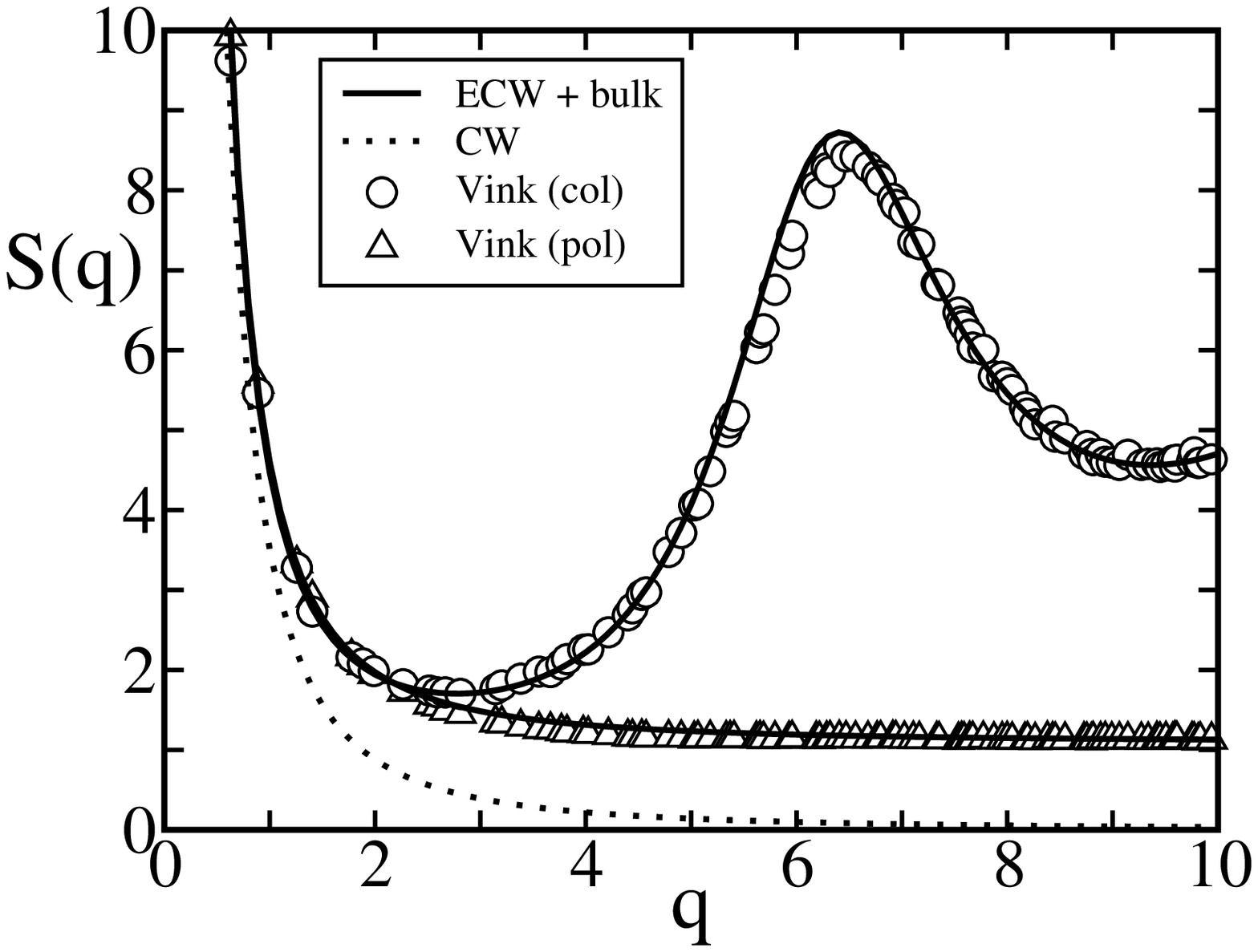}
\caption{MC results by Vink {\em et al.} (Ref.~\onlinecite{Vink}) for the surface structure factor
(in units of $d^4$) versus $q$ (in units of $1/d$) using the colloidal
particles (circles) and polymer particles (triangles) to define $z_G$.
The dotted line is the capillary wave model and the drawn lines are the
combination of the extended capillary wave model and the bulk correlation
function. In this example $\varepsilon\!=\!$ 1.8, $\eta_{p}\!=\!$ 1.0
and $L/W\!=\!$ 3.}
\label{fig:Vink2}
\end{figure}
The excellent agreement between Eq.(\ref{eq:S(q)_ECWB}) and the MC
simulations is even more apparent in Figure \ref{fig:Vink2} where
the results in Figure \ref{fig:Vink1} are redrawn on a linear scale.
In Figure \ref{fig:Vink2} we also show the simulation results \cite{Vink}
and the corresponding fit using the {\em polymer particles} to define
the location of the interface. As the polymer-polymer interactions are
considered ideal, the bulk structure factor $S_b(q) \!=\! 1$ in this case.

It is important to note that, effectively, the inclusion of a bending
rigidity in the capillary wave model results in the presence of an
additive factor in $S(q)$ that is adjusted, see Eq.(\ref{eq:S(q)_ECWB}).
The determination of the value for $k^{ic}$ from the behavior of $S(q)$ near
$q \!=\! 0$, therefore requires one to take into account the presence of
the bulk-like fluctuations since they {\em also} contribute as an additive constant,
${\cal N}_L \, S_b(0)$, near $q \!=\! 0$. This means that even though the
MC simulation results of Vink {\em et al.} \cite{Vink} are very accurately
described by Eq.(\ref{eq:S(q)_ECWB}), the resulting value obtained for $k^{ic}$
sensitively depends on the theoretical expression used for $S_b(0)$.
Here we have simply approximated the bulk correlation function by the
Percus-Yevick hard-sphere expression in the liquid \cite{PY}, but one
could imagine more sophisticated expressions leading to a somewhat
different value for $k^{ic}$.

In the next sections we investigate whether the values for the
bending rigidity obtained from the simulations (Table 1), can also
be described in the context of a molecular theory.

\begin{table}
\centering
\begin{tabular}{c|c|c|c|c|c|}
$\eta_{p}$ & $\eta_{\ell}$ & $\eta_v$ & $\sigma$ & $k$ & $\sqrt{-k/\sigma}$ \\
\hline
0.9 & 0.2970 & 0.0141 & 0.1532 & -0.045 (15) & 0.40 \\
1.0 & 0.3271 & 0.0062 & 0.2848 & -0.07 (2)   & 0.50 \\
1.1 & 0.3485 & 0.0030 & 0.4194 & -0.10 (3)   & 0.49 \\
1.2 & 0.3647 & 0.0018 & 0.5555 & -0.14 (3)   & 0.50 \\
\end{tabular}
\caption{Listed are the simulation results (Ref.~\onlinecite{Vink}) for the polymer volume
fraction $\eta_{p}$, liquid and vapor colloidal volume fractions, $\eta_{\ell}$ and $\eta_v$,
surface tension $\sigma$ (in units of $k_{\rm B} T / d^2$), bending rigidity $k$
(in units of $k_{\rm B} T$; in parenthesis the estimated error in the last digit),
and $\sqrt{-k/\sigma}$ (in units of $d$).}
\end{table}

\section{Density functional theory}

Our task in this section is straightforward. Using the expression
for $\rho(\vec{r})$ given in Eq.(\ref{eq:rho_ECW}), we determine
$\Delta \Omega$ and the resulting $\sigma(q)$. To achieve this, we
need a model for the free energy and define a procedure to determine the
density profiles, $\rho_0(z)$ and $\rho_1(z)$, that are present in
the expression for $\rho(\vec{r})$. We choose to perform these tasks
in the context of density functional theory (DFT).

In the density functional theory for an inhomogeneous system that
we consider \cite{Evans, Henderson, RW, Mecke}, the free energy is given by the
free energy of the reference hard sphere system augmented by an integral,
non-local term that considers the attractive part of the interaction potential,
$U(r)\!=\!U_{\rm hs}(r) + U_{\rm att}(r)$,
\begin{eqnarray}
\label{eq:Omega}
\Omega[\rho] &=& \int \!\! d\vec{r}_1 \; g_{\rm hs}(\rho) \\
&& + \frac{1}{2} \int \!\! d\vec{r}_1 \! \int \!\! d\vec{r}_{2} \;
U_{\rm att}(r) \, \rho(\vec{r}_1) \rho(\vec{r}_2) \,. \nonumber
\end{eqnarray} 
For explicit calculations, $g_{\rm hs}(\rho)$ is taken to be of
the Carnahan-Starling form \cite{CS}:
\begin{equation}
\label{eq:g_hs(rho)}
g_{\rm hs}(\rho) = k_{\rm B} T \, \rho \, \ln(\rho) 
+ k_{\rm B} T \, \rho \, \frac{(4 \eta - 3 \eta^2)}{(1 - \eta)^2} - \mu \rho \,,
\end{equation} 
where $\eta\!\equiv\!(\pi/6) \, \rho \, d^3$. In the uniform bulk
region, the free energy equals
\begin{equation}
\label{eq:g(rho)}
\frac{\Omega(\rho)}{V} \equiv g(\rho) = g_{\rm hs}(\rho) - a \rho^2 \,,
\end{equation} 
with the van der Waals parameter $a$ given by \cite{RW}
\begin{equation}
a \equiv - \frac{1}{2} \, \int \!\! d\vec{r}_{12} \; U_{\rm att}(r) \,.
\end{equation} 
The integration over $\vec{r}_{12}$ is restricted to the
region $r\!>\!d$. This is not explicitly indicated; instead,
we adhere to the convention that the attractive part
of the interaction potential $U_{\rm att}(r)\!=\!0$ when $r\!<\!d$.
The chemical potential $\mu$ is fixed by the condition of two-phase
coexistence, $\mu\!=\!\mu_{\rm coex}$, which implies that $\mu_{\rm coex}$,
$\rho_v$, and $\rho_{\ell}$ are determined from the set of equations:
$g^{\prime}(\rho_v)\!=\!0$, $g^{\prime}(\rho_{\ell})\!=\!0$,
and $g(\rho_v)\!=\!g(\rho_{\ell})\!=\!-p$.

To determine the change in free energy due to density fluctuations,
we insert the expression for $\rho(\vec{r})$ given by Eq.(\ref{eq:rho_ECW})
into the expression for $\Omega$ in Eq.(\ref{eq:Omega}). One then finds for
$\Delta \Omega \!=\! \Omega - \sigma A$
\begin{eqnarray}
\label{eq:Delta_Omega}
&& \Delta \Omega = \frac{1}{8} \int \!\! d\vec{r}_1 \;
\Big\{ g_{\rm hs}^{\prime\prime}(\rho_0) \rho_1(z_1)^2 \left[
\Delta h(\vec{r}_{1,\parallel}) \right]^2 \nonumber \\
&& + \frac{1}{8} \int \!\! d\vec{r}_1 \int \!\! d\vec{r}_{12} \; U_{\rm att}(r) \times \nonumber \\
&& \left\{ - 2 \, \rho_0^{\prime}(z_1) \rho_0^{\prime}(z_2)
\left[ h(\vec{r}_{2,\parallel}) - h(\vec{r}_{1,\parallel}) \right]^2 \right. \nonumber \\
&& \hspace*{7pt} + 4 \, \rho_1(z_1) \rho_0^{\prime}(z_2) \, \Delta h(\vec{r}_{1,\parallel})
\left[ h(\vec{r}_{2,\parallel}) - h(\vec{r}_{1,\parallel}) \right] \nonumber \\
&& \hspace*{7pt} + \rho_1(z_1) \rho_1(z_2) \,
\Delta h(\vec{r}_{1,\parallel}) \Delta h(\vec{r}_{2,\parallel}) \Big\} \,.
\end{eqnarray}
Even though the derivation is somewhat different, this expression equals that
given by Mecke and Dietrich \cite{Mecke} apart from a gravity term that
was included in their expression. To cast $\Delta \Omega$ in the form
of Eq.(\ref{eq:Omega_ECW}), we take the Fourier Transform. One then finds
for $\sigma(q)$ \cite{Mecke}
\begin{eqnarray}
\label{eq:sigma(q)}
\sigma(q) &=& \int\limits_{-\infty}^{\infty} \!\!\! dz_1 \! \int\limits_{-\infty}^{\infty} \!\!\! dz_{12}
\; \left[ \frac{\omega(q,z_{12}) - \omega_0(z_{12})}{q^2} \right] \nonumber \\
&& \!\! \times \left[ \, \rho_0^{\prime}(z_1) \rho_0^{\prime}(z_2)
- q^2 \, \rho_1(z_1) \rho_0^{\prime}(z_2) \, \right] \nonumber \\
&& \!\! + \frac{q^2}{4} \int\limits_{-\infty}^{\infty} \!\!\! dz_1 \!\! \int\limits_{-\infty}^{\infty} \!\!\! dz_{12} \;
\omega(q,z_{12}) \, \rho_1(z_1) \rho_1(z_2) \nonumber \\
&& \!\! + \frac{q^2}{4} \int\limits_{-\infty}^{\infty} \!\!\! dz_1 \; g_{\rm hs}^{\prime\prime}(\rho_0) \rho_1(z_1)^2 \,.
\end{eqnarray}
Here we have defined the (parallel) Fourier Transform of the interaction
potential
\begin{eqnarray}
\label{eq:w(q,z12)}
\omega(q,z_{12}) &\equiv& \int \!\! d\vec{r}_{\parallel} \,
e^{- i \vec{q} \cdot \vec{r}_{\parallel}} \, U_{\rm att}(r) \\
&=& 2 \pi \! \int\limits_{0}^{\infty} \!\! dr_{\parallel} \, r_{\parallel} \,
J_0(qr_{\parallel}) \, U_{\rm att}(r) \,.  \nonumber
\end{eqnarray}

As a first step, we determine the leading contribution to $\sigma(q)$ given
by the surface tension of the {\em planar} interface, $\sigma\!=\!\sigma(q\!=\!0)$.
Then, one needs to consider the two leading contributions
in the expansion of $\omega(q,z_{12})$ in $q^2$:
\begin{equation}
\omega(q,z_{12}) = \omega_0(z_{12}) + \omega_2(z_{12}) {q^2} + \ldots \,,
\end{equation}
where
\begin{eqnarray}
\omega_0(z_{12}) &\equiv& \int \!\! d\vec{r}_{\parallel} \; U_{\rm att}(r) \,, \nonumber \\
\omega_2(z_{12}) &\equiv& - \frac{1}{4} \int \!\! d\vec{r}_{\parallel} \; r_{\parallel}^2 \, U_{\rm att}(r) \,.
\end{eqnarray} 
The surface tension thus becomes
\begin{equation}
\label{eq:sigma}
\sigma = \int\limits_{-\infty}^{\infty} \!\!\! dz_1 \int\limits_{-\infty}^{\infty} \!\!\! dz_{12} \;
\omega_2(z_{12}) \, \rho_0^{\prime}(z_1) \rho_0^{\prime}(z_2) \,.
\end{equation}
The (planar) density profile $\rho_0(z)$, featured in the above expression
for $\sigma$, is determined from minimizing the free energy functional
$\Omega[\rho]$ in Eq.(\ref{eq:Omega}) in planar symmetry. The Euler-Lagrange
equation that minimizes $\Omega[\rho]$ is then given by:
\begin{eqnarray}
\label{eq:EL_0}
g^{\prime}_{\rm hs}(\rho_0) &=& - \int \!\! d\vec{r}_{12} \; U_{\rm att}(r) \, \rho_0(z_2) \nonumber \\
&=& - \int\limits_{-\infty}^{\infty} \!\!\! dz_{12} \; \omega_0(z_{12}) \, \rho_0(z_2) \,,
\end{eqnarray}
which can be solved explicitly to obtain $\rho_0(z)$ and thus $\sigma$.

The evaluation of further contributions to $\sigma(q)$ requires one to
determine the density profile $\rho_1(z)$. Just like $\rho_0(z)$, one would
like to determine the density profile $\rho_1(z)$ from a minimization procedure.
One then has to determine the energetically most favorable density profile
for a {\em given} curvature of the surface \cite{Parry94}. This turns out to be
not so straightforward, since one then has to specify in what way the curvature
is set to its given value. Several approaches have been suggested, which we
shall now discuss.

\begin{itemize}

\item {\bf Mecke and Dietrich approach}. In this approach a certain
form for $\rho_1(z)$ is directly hypothesized \cite{Mecke}:
\begin{equation}
\label{eq:rho1_MD}
\rho^{\rm MD}_1(z) = - \frac{C_{\rm H}}{2 \pi} \, \Delta \rho \, \xi \, f_{\rm H}(z/\xi) \,,
\end{equation}
with $\xi$ the bulk correlation length and $f_{\rm H}(x)\!\equiv\!x \sinh(x/2)/\cosh^2(x/2)$.
The coefficient $C_{\rm H}$ in this expression can be used as a fit parameter.
This practical approach is certainly legitimate, but one would like to
also be able to formulate a molecular basis for this expression. 

\item {\bf Equilibrium approach}. Rather than the surface being curved by
surface fluctuations, in this approach the interface is curved by changing
the value of the chemical potential to a value {\em off-coexistence}. One
then considers the density profile of a spherically or cylindrically shaped
liquid droplet in metastable equilibrium with a bulk vapor \cite{Blokhuis93, Gompper}.
This approach is equivalent to adding an external field to the free energy
\begin{eqnarray}
\Omega^{\prime}[\rho] &=& \Omega[\rho] + \int \!\! d\vec{r} \; V_{\rm ext}(\vec{r}) \, \rho(\vec{r}) \nonumber \\
&=& \Omega[\rho] - \int \!\! d\vec{r} \; \Delta \mu \, \rho(\vec{r}) \,,
\end{eqnarray}
where $\Delta\mu\!=\!\mu-\mu_{\rm coex}$. The downside of the equilibrium
approach is that the external field $V_{\rm ext}(\vec{r})\!=\!-\Delta \mu$
is uniform throughout the system and thus also affects the bulk densities
far from the interfacial region. This seems inappropriate for the description
of the density fluctuations considered here since we have that $\mu\!=\!\mu_{\rm coex}$
and the bulk densities are unaltered by the curvature of the surface fluctuations.

\item {\bf Local external field}. In this approach, one again adds to
the free energy an external field, but, to ensure that the bulk regions
are unaffected, one assumes that it is peaked infinitely sharply at
$z\!=\!0$ \cite{Parry94, Blokhuis99}:
\begin{equation}
V_{\rm ext}(\vec{r}) = \lambda \, \delta(z) \, \Delta h(\vec{r}_{\parallel}) \,.
\end{equation}
In this case, the external field only acts as a Lagrange multiplier in the
minimization procedure to ensure that the curvature $\Delta h(\vec{r}_{\parallel})$
is set to a certain value; it is not included in the expression for the
free energy. The downside of this method is that the resulting density
profile $\rho_1(z)$ has a discontinuous first derivative at $z\!=\!0$,
which is, from a physical point of view, not so appealing \cite{FJ}.
Furthermore, the discontinuous nature of $\rho_1(z)$ prohibits an analytical
simplification using a gradient expansion.

\end{itemize}

\noindent
In the present approach, we suggest to add an external field acting as a
Lagrange multiplier that is unequal to zero only in the interfacial
region (the bulk densities are unaffected), but which is not infinitely
sharp-peaked. It seems natural to choose a peak-width of the order of
the thickness of the interfacial region. It thus seems convenient
to choose $V_{\rm ext}(\vec{r}) \! \propto \! \rho^{\prime}_0(z)$:
\begin{equation}
\label{eq:V_ext}
V_{\rm ext}(\vec{r}) = \lambda \, \rho^{\prime}_0(z) \, \Delta h(\vec{r}_{\parallel}) \,.
\end{equation}
This choice for $V_{\rm ext}(\vec{r})$ constitutes our fundamental `Ansatz'
for the determination of $\rho_1(z)$. The Lagrange multiplier $\lambda$
is not a free parameter but set by the imposed curvature, as demonstrated below.

The addition of an external field to the free energy results in the following
Euler-Lagrange equation:
\begin{equation}
\label{eq:EL}
g^{\prime}_{\rm hs}(\rho) = - \int \!\! d\vec{r}_{12} \; U_{\rm att}(r) \, \rho(\vec{r}_2)
- V_{\rm ext}(\vec{r})\,.
\end{equation}
Using the external field given in Eq.(\ref{eq:V_ext}), we insert the fluctuating
density given by Eq.(\ref{eq:rho_ECW}) into the above Euler-Lagrange equation.
In order for the resulting equation to hold independently of the value of
$h(\vec{r}_{\parallel})$ or $\Delta h(\vec{r}_{\parallel})$, one finds,
besides Eq.(\ref{eq:EL_0}), the following equation to determine $\rho_1(z)$:
\begin{eqnarray}
\label{eq:EL_1}
&& g^{\prime\prime}_{\rm hs}(\rho_0) \rho_1(z_1) = 
- \int\limits_{-\infty}^{\infty} \!\!\! dz_{12} \; \omega_0(z_{12}) \, \rho_1(z_2) \nonumber \\
&& + 2 \int\limits_{-\infty}^{\infty} \!\!\! dz_{12} \; \omega_2(z_{12}) \, \rho^{\prime}_0(z_2)
+ 2 \, \lambda \, \rho^{\prime}_0(z_1) \,.
\end{eqnarray}
The value of the Lagrange multiplier can be determined by multiplying both
sides of the above expression by $\rho^{\prime}_0(z_1)$ and integrating over $z_1$:
\begin{equation}
\lambda = - \sigma / \left[ \int \!\! dz \; [ \rho_0^{\prime}(z) ]^2 \right] \,.
\end{equation}
One may now verify that if $\rho_1(z)$ is a particular solution of
Eq.(\ref{eq:EL_1}) that then also $\rho_1(z) + \alpha \, \rho^{\prime}_0(z)$
is a solution on account of Eq.(\ref{eq:EL_0}).

It is convenient to use the Euler-Lagrange equation in Eq.(\ref{eq:EL_1})
to remove the explicit appearance of $g^{\prime\prime}_{\rm hs}(\rho_0)$
in the expression for $\sigma(q)$ in Eq.(\ref{eq:sigma(q)}).
The resulting $\sigma(q)$ is written as the sum of a term that
depends only on the density profile $\rho_0(z)$ and one term
that also depends on the density profile $\rho_1(z)$
\begin{equation}
\label{eq:sigma_tot}
\sigma(q) = \sigma_0(q) + k_1 \, q^2 + {\cal O}(q^4) \,,
\end{equation}
with 
\begin{eqnarray}
\label{eq:sigma_0(q)_k1}
\sigma_0(q) &\equiv& \int\limits_{-\infty}^{\infty} \!\!\! dz_1 \! \int\limits_{-\infty}^{\infty} \!\!\! dz_{12}
\left[ \frac{\omega(q,z_{12}) - \omega_0(z_{12})}{q^2} \right] \nonumber \\
&& \hspace*{60pt} \times \, \rho_0^{\prime}(z_1) \rho_0^{\prime}(z_2) \,, \nonumber \\
k_1 &\equiv&  - \frac{1}{2} \int\limits_{-\infty}^{\infty} \!\!\! dz_1 \! \int\limits_{-\infty}^{\infty} \!\!\! dz_{12} \;
\omega_2(z_{12}) \, \rho_1(z_1) \rho_0^{\prime}(z_2) \nonumber \\
&& + \frac{\lambda}{2} \int\limits_{-\infty}^{\infty} \!\!\! dz_1 \; \rho_1(z_1) \rho_0^{\prime}(z_1) \,.
\end{eqnarray}
With the above expression for $k_1$ it is now also possible to verify
that when the density profile $\rho_1(z)$ is shifted by a factor
$\alpha \rho^{\prime}_0(z)$, that the resulting effect on the bending
parameters is that given by Eq.(\ref{eq:k_shift}).

The procedure to determine $\sigma(q)$, and therefore $\sigma$, $k$,
$k_s$, and $\ell_k$, is now as follows: assuming a certain form for
the attractive part of the interaction potential, $\rho_0(z)$ is
obtained from solving Eq.(\ref{eq:EL_0}), which is then
inserted into Eq.(\ref{eq:EL_1}) to solve for $\rho_1(z)$ explicitly.
The two density profiles thus obtained are inserted into
Eq.(\ref{eq:sigma_0(q)_k1}) to yield $\sigma_0(q)$ and $k_1$.
This procedure is carried out in the next two sections considering
short-ranged forces and long-ranged forces ($U(r)\!\propto\!1/r^6$).
In general the density profiles $\rho_0(z)$ and $\rho_1(z)$
need to be determined numerically. We shall, however, also provide
an approximation scheme, based on the gradient expansion, that is exact
near the critical point, but which also gives an excellent approximation
far from it.

\subsection{Gradient expansion}

The gradient approximation \cite{RW} is based on the assumption that
the spatial variation of the density profile is small, i.e.
\begin{equation}
\label{eq:gradient}
\rho(z_2) = \rho(z_1) + z_{12} \, \rho^{\prime}(z_1)
+ \frac{z_{12}^2}{2} \, \rho^{\prime\prime}(z_1) + \ldots
\end{equation}
In the gradient expansion, the Euler-Lagrange equation in Eq.(\ref{eq:EL_0})
for $\rho_0(z)$ reduces to
\begin{equation}
\label{eq:EL_0_SQ}
g^{\prime}(\rho_0) = 2 m \, \rho^{\prime\prime}_0(z) \,,
\end{equation}
where $m$ is the van der Waals squared-gradient coefficient \cite{RW}
\begin{eqnarray}
m &\equiv& - \frac{1}{12} \, \int \!\! d\vec{r}_{12} \; r^2 \, U_{\rm att}(r) \\
&=& \frac{1}{2} \int\limits_{-\infty}^{\infty} \!\!\! dz_{12} \; \omega_2(z_{12})
= \frac{1}{4} \int\limits_{-\infty}^{\infty} \!\!\! dz_{12} \; z_{12}^2 \, \omega_0(z_{12}) \,. \nonumber
\end{eqnarray}
In the gradient expansion, the Euler-Lagrange equation in Eq.(\ref{eq:EL_1})
for $\rho_1(z)$ reduces to
\begin{eqnarray}
\label{eq:EL_1_SQ}
&& m \, \frac{\rho_0^{\prime\prime\prime}(z_1)}{\rho_0^{\prime}(z_1)} \rho_1(z_1)
= m \, \rho_1^{\prime\prime}(z_1) \\
&& \hspace*{40pt} + \int\limits_{-\infty}^{\infty} \!\!\! dz_{12} \; \omega_2(z_{12}) \,
\rho_0^{\prime}(z_2) + \lambda \, \rho^{\prime}_0(z_1) \,, \nonumber
\end{eqnarray}
where we have used Eq.(\ref{eq:EL_0_SQ}) to replace $g^{\prime\prime}_{\rm hs}(\rho_0)$.

First, we consider the determination of the density profile $\rho_0(z)$.
The gradient expansion becomes exact near the critical point
where $g(\rho)$ takes on the usual double-well form
\begin{equation}
\label{eq:rho^4}
g(\rho) + p = \frac{m}{(\Delta \rho)^2 \, \xi^2} \, (\rho - \rho_{\ell})^2 \, (\rho - \rho_v)^2 \,.
\end{equation}
Using this form for $g(\rho)$, the solution of the Euler-Lagrange equation in
Eq.(\ref{eq:EL_0_SQ}) gives the usual $\tanh$-form for $\rho_0(z)$ \cite{RW}:
\begin{equation}
\label{eq:tanh}
\rho_0(z) = \frac{1}{2} (\rho_{\ell} + \rho_v) - \frac{\Delta \rho}{2} \, \tanh(z/2\xi) \,,
\end{equation}
with the bulk correlation length $\xi$ a measure of the interfacial thickness.

Even though the $\tanh$-form for the density profile $\rho_0(z)$ is
derived assuming proximity to the critical point, it turns out that
it also provides a good approximation away from it when one determines
the value of $\xi$ by fitting the surface tension to its form near the
critical point. In the squared-gradient approximation, Eq.(\ref{eq:sigma})
reduces to the familiar expression \cite{RW}
\begin{eqnarray}
\label{eq:sigma_SQ}
\sigma &=& 2 \, m  \int\limits_{-\infty}^{\infty} \!\!\! dz \; \rho_0^{\prime}(z)^2 \nonumber \\
&=& 2 \, \sqrt{m}  \int\limits_{\rho_v}^{\rho_{\ell}} \!\! d\rho \; \sqrt{g(\rho) + p} \,.
\end{eqnarray}
On the one hand, the surface tension can be determined from the above
approximation using the full form for $g(\rho)$ given in Eqs.(\ref{eq:g_hs(rho)})
and (\ref{eq:g(rho)}):
\begin{eqnarray}
\label{eq:sigma_SQ_full}
\sigma &=& 2 \, \sqrt{m}  \int\limits_{\rho_v}^{\rho_{\ell}} \!\! d\rho  \; 
\bigl[ k_{\rm B} T \, \rho \, \ln(\rho) \\
&& + k_{\rm B} T \, \rho \, \frac{(4 \eta - 3 \eta^2)}{(1 - \eta)^2}
- \mu_{\rm coex} \rho - a \rho^2 + p \bigr]^{\!\frac{1}{2}} \,. \nonumber
\end{eqnarray}
On the other hand, near the critical point $g(\rho)$ takes on the double-well
form in Eq.(\ref{eq:rho^4}) and $\sigma$ is calculated as
\begin{equation}
\label{eq:sigma_SQ_cp}
\sigma = \frac{m \, (\Delta \rho)^2}{3 \, \xi} \,.
\end{equation}
Now, we define the value of $\xi$ such that the two expressions for the
surface tension in Eqs.(\ref{eq:sigma_SQ_full}) and (\ref{eq:sigma_SQ_cp})
are equal. This gives for $\xi$:
\begin{equation}
\label{eq:xi}
\xi \equiv \frac{m \, (\Delta \rho)^2}{3 \, \sigma} \,,
\end{equation}
with $\sigma$ given by Eq.(\ref{eq:sigma_SQ_full}).

Next, we turn to the evaluation of $\rho_1(z)$ from Eq.(\ref{eq:EL_1_SQ}).
This requires one to make a distinction between short-ranged forces and
long-ranged forces.

\section{DFT: short-ranged interactions}

Although the analysis below is quite generally valid for all short-ranged
interaction potentials, whenever we show explicit results, we consider for
$U_{\rm att}(r)$ the Asakura-Oosawa-Vrij depletion interaction potential
$U_{\rm dep}(r)$ as an example \cite{AOVrij}:
\begin{equation}
U_{\rm dep}(r) = \frac{- k_{\rm B} T \, \eta_p}{2 \, (\varepsilon-1)^3}
\left[ \, 2 \, \varepsilon^3 - 3 \, \varepsilon^2 \left( \frac{r}{d} \right)
+ \left( \frac{r}{d} \right)^{\!3} \, \right]
\end{equation}
where the intermolecular distance is in the range $1\!<\!r/d\!<\!\varepsilon$.
Interaction parameters based on the depletion potential are listed in the
Appendix. The strength of the depletion interaction potential as determined
by the polymer volume fraction $\eta_p$ determines the location in the 
phase diagram \cite{Gast, Lekkerkerker90}; for comparison with other results,
it is, however, more convenient to use the colloidal density difference
$\Delta \eta \!\equiv\! \eta_{\ell}-\eta_v$ as thermodynamic variable
\cite{Kuipers}.

\begin{figure}
\centering
\includegraphics[angle=270,width=200pt]{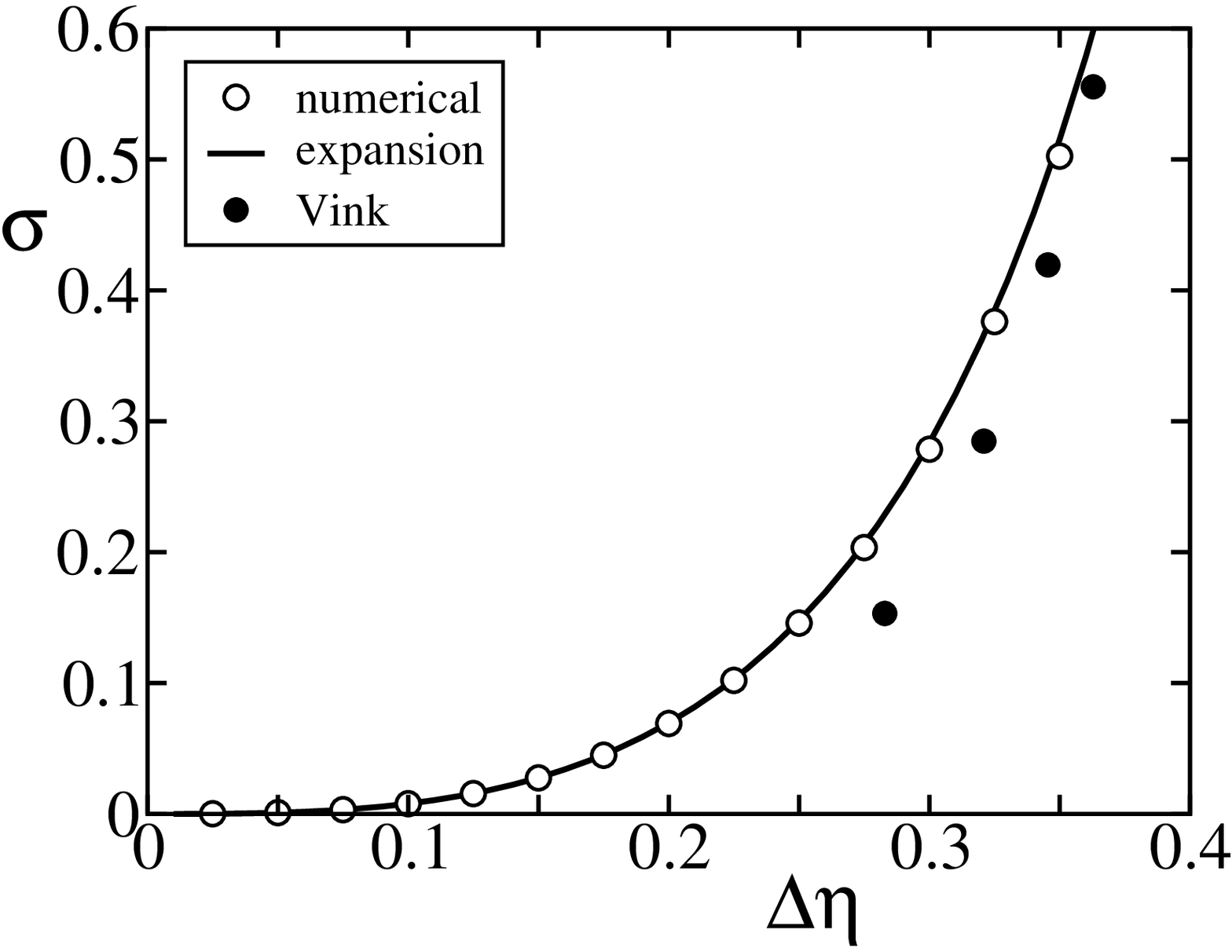}
\caption{Surface tension in units of $k_{\rm B} T / d^2$
versus the volume fraction difference, $\Delta \eta$.
In this example $\varepsilon\!=\!1.8$; symbols are numerical results,
the drawn line is the gradient expansion approximation, filled symbols
are results from the MC simulations by Vink {\em et al.} (Ref.~\onlinecite{Vink}).}
\label{fig:sigma}
\end{figure}
In Figure \ref{fig:sigma}, the surface tension is shown as a function of $\Delta \eta$.
The open circles are obtained from numerically solving the Euler-Lagrange
equation in Eq.(\ref{eq:EL_0}) for $\rho_0(z)$ and inserting the result
into Eq.(\ref{eq:sigma}). The drawn line is the gradient expansion approximation
for $\sigma$ in Eq.(\ref{eq:sigma_SQ_full}). Also shown are results from the MC
simulations by Vink {\em et al.} \cite{Vink}. The gradient expansion gives a very
good approximation to the numerical results and is in good agreement with the
simulations.

For short-ranged forces the expansion in $q^2$ of the expression for $\omega(q,z_{12})$
as defined by Eq.(\ref{eq:w(q,z12)}) can be continued to ${\cal O}(q^4)$:
\begin{equation}
\omega(q,z_{12}) = \omega_0(z_{12}) + \omega_2(z_{12}) \, q^2 + \omega_4(z_{12}) \, q^4 + \ldots
\end{equation}
where
\begin{equation}
\label{eq:w4(z12)}
\omega_4(z_{12}) \equiv \frac{1}{64} \int \!\! d\vec{r}_{\parallel} \; r_{\parallel}^4 \, U_{\rm att}(r) \,.
\end{equation} 
With this expansion, $\sigma(q)$ in Eq.(\ref{eq:sigma_0(q)_k1}) can now be
written in the form of Eq.(\ref{eq:sigma_ECW_SR})
\begin{eqnarray}
\label{eq:sigma(q)_SR}
\sigma(q) &=& \sigma + k \, q^2 + {\cal O}(q^4) \nonumber \\
&=& \sigma + k_0 \, q^2 + k_1 \, q^2 + {\cal O}(q^4) \,,
\end{eqnarray}
with the bending rigidity $k \!=\! k_0 + k_1$ and  
\begin{eqnarray}
\label{eq:k_SR}
k_0 &=& \int\limits_{-\infty}^{\infty} \!\!\! dz_1 \! \int\limits_{-\infty}^{\infty} \!\!\! dz_{12} \;
\omega_4(z_{12}) \, \rho_0^{\prime}(z_1) \rho_0^{\prime}(z_2) \,, \nonumber \\
k_1 &=& - \frac{1}{2} \int\limits_{-\infty}^{\infty} \!\!\! dz_1 \! \int\limits_{-\infty}^{\infty} \!\!\! dz_{12} \;
\omega_2(z_{12}) \, \rho_1(z_1) \rho_0^{\prime}(z_2) \nonumber \\
&& + \frac{\lambda}{2} \int\limits_{-\infty}^{\infty} \!\!\! dz_1 \; \rho_1(z_1) \rho_0^{\prime}(z_1) \,.
\end{eqnarray}
Next, we proceed to evaluate these expressions in the gradient expansion.

\subsection{Gradient expansion for short-ranged forces}

In the gradient expansion, $k_0$ and $k_1$ in Eq.(\ref{eq:k_SR})
reduce to:
\begin{eqnarray}
\label{eq:k0k1_SQ_SR}
k_0 &=& - \frac{B}{2} \int\limits_{-\infty}^{\infty} \!\!\! dz \; \rho_0^{\prime}(z)^2 \,, \nonumber \\
k_1 &=& - 2 \, m \int\limits_{-\infty}^{\infty} \!\!\! dz \; \rho_1(z) \, \rho_0^{\prime}(z) \,,
\end{eqnarray}
where we have used the fact that to leading order in the gradient
expansion $\lambda\!\approx\!-2m$, and where we have defined \cite{Varea}
\begin{eqnarray}
B &\equiv& - \frac{1}{60} \, \int \!\! d\vec{r}_{12} \; r^4 \, U_{\rm att}(r) \\
&=& -2 \! \int\limits_{-\infty}^{\infty} \!\!\! dz_{12} \; \omega_4(z_{12})
= \frac{1}{2} \! \int\limits_{-\infty}^{\infty} \!\!\! dz_{12} \; z_{12}^2 \, \omega_2(z_{12}) \,. \nonumber
\end{eqnarray}
Inserting the $\tanh$-form for $\rho_0(z)$ into Eq.(\ref{eq:k0k1_SQ_SR}),
one directly obtains for $k_0$
\begin{equation}
\label{eq:k0_SQ_SR}
k_0 = - \frac{B \, (\Delta \rho)^2}{12 \, \xi} = - \frac{B \, \sigma}{4 \, m} \,,
\end{equation}
where we have used the expression for $\xi$ in Eq.(\ref{eq:xi}) to rewrite $k_0$
as the latter expression. 

To evaluate $k_1$ in Eq.(\ref{eq:k0k1_SQ_SR}), we need to determine
$\rho_1(z)$ from the Euler-Lagrange equation in Eq.(\ref{eq:EL_1_SQ}).
For short-ranged forces, Eq.(\ref{eq:EL_1_SQ}) reduces to
\begin{equation}
\label{eq:EL_1_SQ_SR}
m \, \frac{\rho_0^{\prime\prime\prime}(z)}{\rho_0^{\prime}(z)} \rho_1(z)
= m \, \rho_1^{\prime\prime}(z) + B \, \rho^{\prime\prime\prime}_0(z) + \beta \, B \, \rho^{\prime}_0(z) \,,
\end{equation}
where we have defined
\begin{equation}
\beta \equiv \frac{\int \! dz \; \rho_0^{\prime\prime}(z)^2}{\int \! dz \; \rho_0^{\prime}(z)^2} \,.
\end{equation}
Using the $\tanh$-profile for $\rho_0(z)$ in Eq.(\ref{eq:tanh}), one has
$\beta\!=\!1/(5 \xi^2)$ and finds for $\rho_1(z)$ from solving the
differential equation in Eq.(\ref{eq:EL_1_SQ_SR}):
\begin{equation}
\label{eq:rho1_cc}
\rho^{cc}_1(z) = - \frac{3 \, B}{10 \, m} \frac{\Delta \rho}{\xi} \,
\frac{\ln(\cosh(z/2\xi))}{\cosh^2(z/2\xi)} \,.
\end{equation}
The above profile corresponds to that obtained using the crossing constraint.
The profile corresponding to the integral constraint follows from
$\rho^{ic}_1\!=\!\rho^{cc}_1 + \alpha \, \rho^{\prime}_0$ (Eq.(\ref{eq:relation_rho1_ic_cc})),
with $\alpha$ determined by Eq.(\ref{eq:alpha}). This gives
\begin{equation}
\label{eq:rho1_ic}
\!\!\! \rho^{ic}_1(z) = \frac{3 \, B}{10 \, m} \frac{\Delta \rho}{\xi} \,
\frac{\left[ \, 1 - \ln(2 \, \cosh(z/2\xi)) \, \right]}{\cosh^2(z/2\xi)} \,.
\end{equation}
\begin{figure}
\centering
\includegraphics[angle=270,width=200pt]{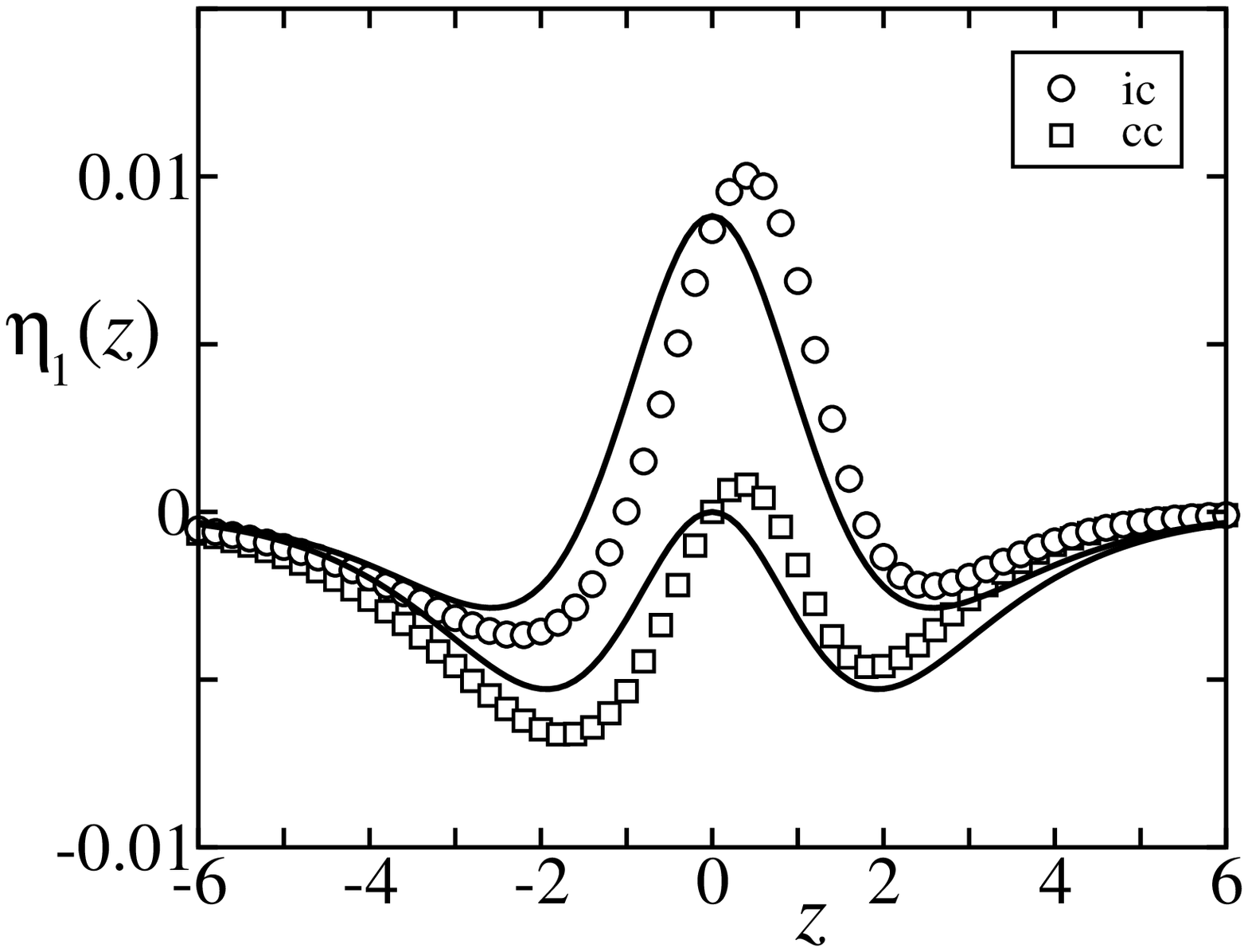}
\caption{Curvature correction to the volume fraction profile
$\eta_1(z)$ as a function of $z$ (in units of $d$) using the
integral constraint (circles) and crossing constraint (squares).
In this example $\Delta \eta\!=\!$ 0.25, $\varepsilon\!=\!1.8$;
symbols are numerical results, the drawn lines are the analytical
profiles from the gradient expansion.}
\label{fig:prof}
\end{figure}
\noindent
In Figure \ref{fig:prof}, typical volume fraction profiles
$\eta_1(z)\!=\!(\pi/6) \, d^3 \, \rho_1(z)$ are shown
for the crossing constraint and the integral constraint. The symbols
are the profiles obtained from numerically solving the Euler-Lagrange
equation in Eq.(\ref{eq:EL_1}), whereas the drawn lines are the approximate
profiles in Eqs.(\ref{eq:rho1_cc}) and (\ref{eq:rho1_ic}) obtained
from the gradient expansion.

Inserting the density profiles $\rho^{ic}_1(z)$ and $\rho^{cc}_1(z)$
into the expression for $k_1$ in Eq.(\ref{eq:k0k1_SQ_SR}), one obtains
\begin{eqnarray}
\label{eq:k1_SQ_SR}
k_1^{cc} &=& - \frac{B \, (\Delta \rho)^2}{\xi} \, \left[ \frac{1}{3} - \frac{2}{5} \, \ln(2) \right] \nonumber \\
&=& - \frac{B \, \sigma}{m} \, \left[ 1 - \frac{6}{5} \, \ln(2) \right] \,, \nonumber \\
k_1^{ic} &=& \frac{B \, (\Delta \rho)^2}{15 \, \xi}  = \frac{B \, \sigma}{5 \, m} \,.
\end{eqnarray}
\begin{figure}
\centering
\includegraphics[angle=270,width=200pt]{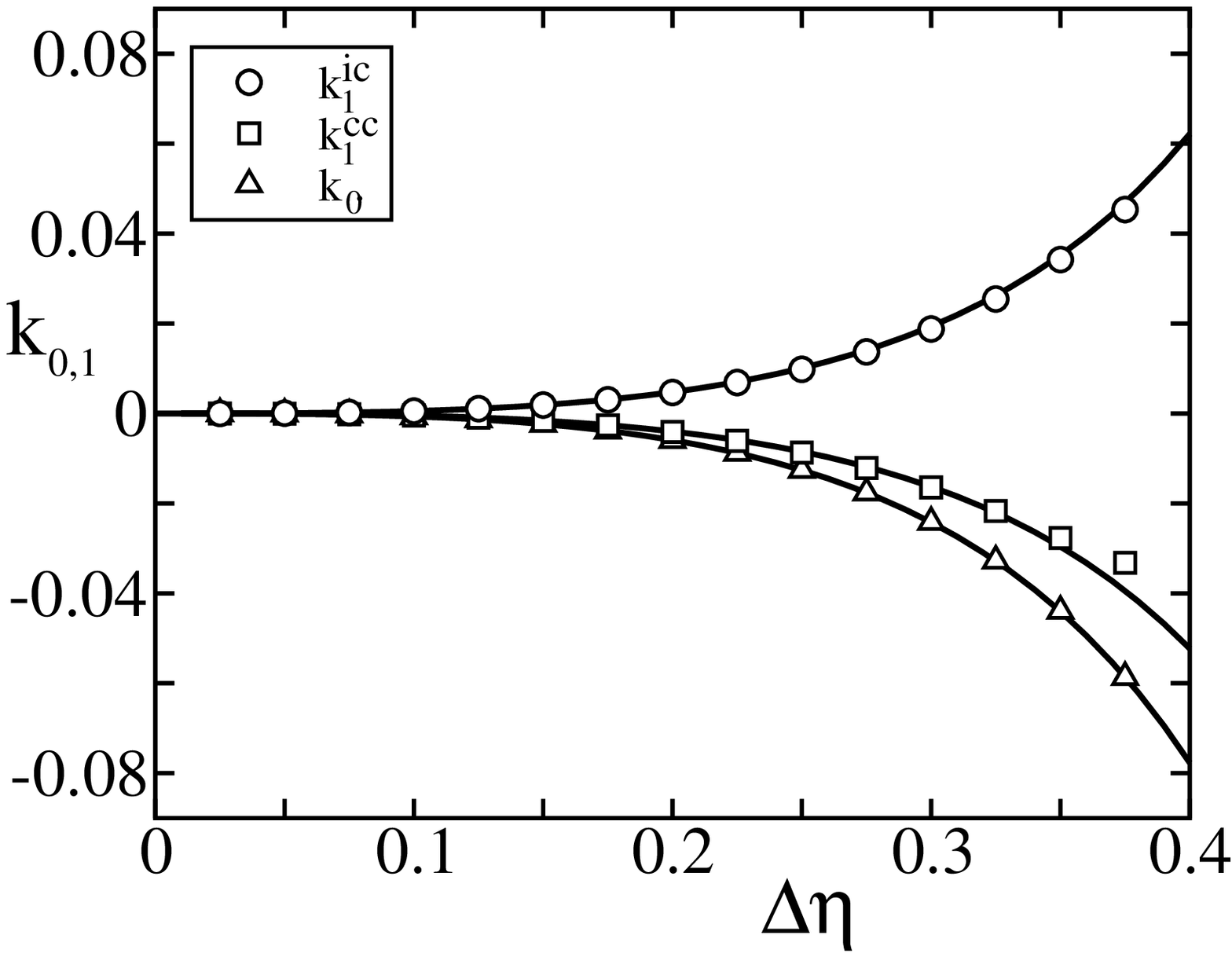}
\caption{Contributions to the bending rigidity $k_0$ and $k_1$ in
units of $k_{\rm B} T$ versus the volume fraction difference, $\Delta \eta$.
In this example $\varepsilon\!=\!1.8$; symbols are numerical results,
the drawn lines are the gradient expansion approximation.}
\label{fig:k_SR1}
\end{figure}
\noindent
In Figure \ref{fig:k_SR1}, $k_0$, $k_1^{ic}$ and $k_1^{cc}$ are shown as a function
of $\Delta \eta$. The open symbols are obtained from numerically solving
Eqs.(\ref{eq:EL_0}) and (\ref{eq:EL_1}) to obtain $\rho_0(z)$ and $\rho_1(z)$
and inserting the result into Eq.(\ref{eq:k_SR}). The drawn lines are the
gradient expansion approximation for $k_0$ in Eq.(\ref{eq:k0_SQ_SR})
and $k_1$ in Eq.(\ref{eq:k1_SQ_SR}).
Adding the results for $k_1$ in Eq.(\ref{eq:k1_SQ_SR}) to $k_0$ in
Eq.(\ref{eq:k0_SQ_SR}), one obtains for the bending rigidities
\begin{eqnarray}
\label{eq:k_SQ_SR}
k^{cc} &=& - \frac{B \, (\Delta \rho)^2}{\xi} \, \left[ \frac{5}{12} - \frac{2}{5} \, \ln(2) \right] \nonumber \\
&=& - \frac{B \, \sigma}{m} \, \left[ \frac{5}{4} - \frac{6}{5} \, \ln(2) \right] \,, \nonumber \\
k^{ic} &=& - \frac{B \, (\Delta \rho)^2}{60 \, \xi} = - \frac{B \, \sigma}{20 \, m} \,.
\end{eqnarray}
\begin{figure}
\centering
\includegraphics[angle=270,width=200pt]{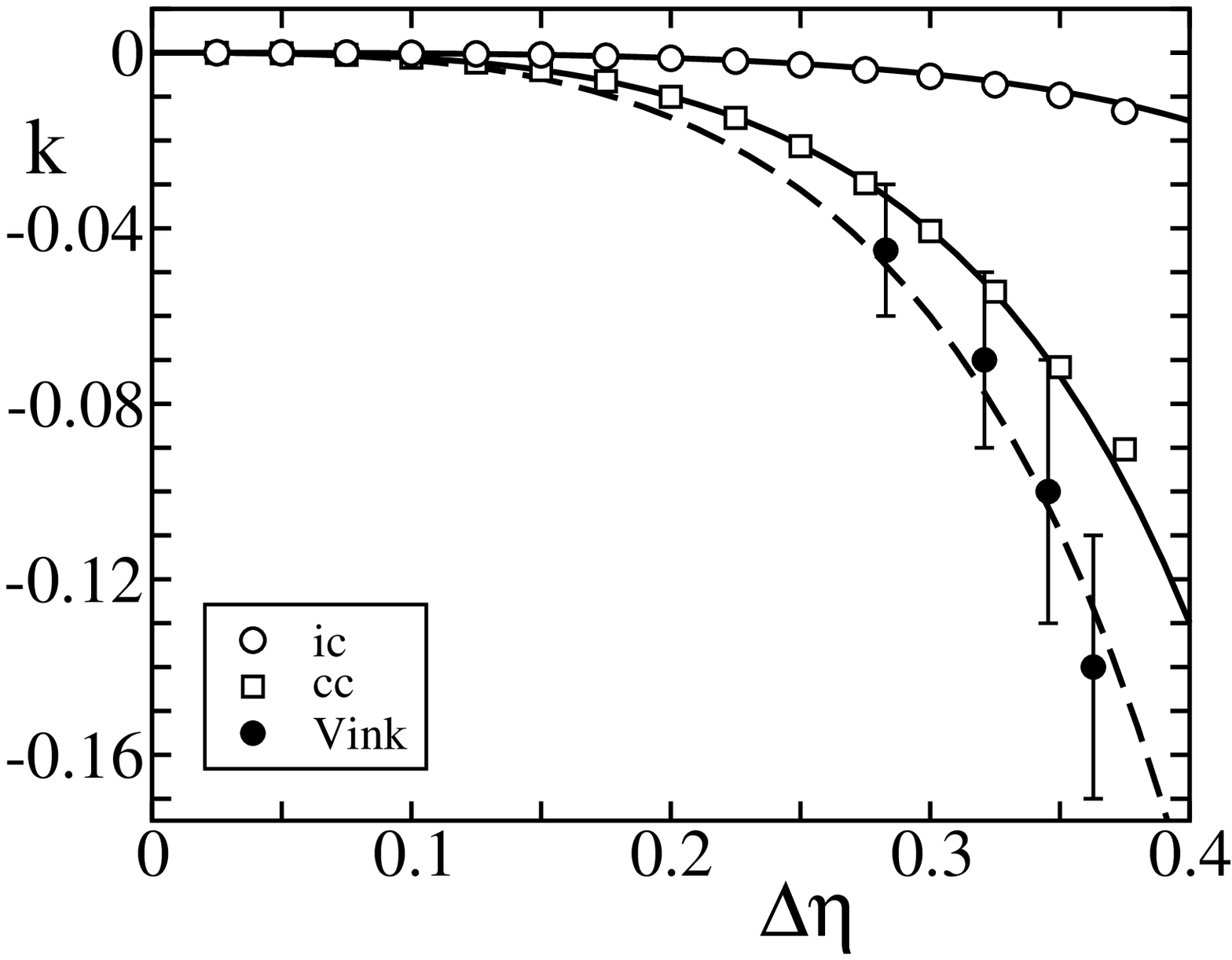}
\caption{Bending rigidity in units of $k_{\rm B} T$
versus the volume fraction difference, $\Delta \eta$, using the
integral constraint (circles) and crossing constraint (squares).
In this example $\varepsilon\!=\!1.8$; open symbols are numerical results,
the drawn lines are the gradient expansion approximation, filled circles
are the MC results by Vink {\em et al.} (Ref.~\onlinecite{Vink}); the dashed line is
the fit $\sqrt{-k/\sigma} \approx$~0.47~$d$.}
\label{fig:k_SR2}
\end{figure}
\noindent
In Figure \ref{fig:k_SR2}, the bending rigidity is shown as a function of
$\Delta \eta$. The open symbols are the numerical results. The drawn lines
are the gradient expansion approximations in Eq.(\ref{eq:k_SQ_SR}),
$\sqrt{-k^{cc}/\sigma} \approx$~0.378~$d$ and
$\sqrt{-k^{ic}/\sigma} \approx$~0.131~$d$.

The resulting values for both bending rigidities, $k^{cc}$ and $k^{ic}$,
are {\em negative} in line with the simulation results of Vink (filled
circles). However, the value of $k^{ic}$, which is the relevant value
when we compare with the simulations, is significantly less negative.
As stressed earlier, the bending rigidity depends on the constraint used to
define the height profile $h(\vec{r}_{\parallel})$ through the contribution
to $k$ coming from $k_1$. We demonstrated that in the crossing constraint $k^{cc}_1$
is negative whereas in the integral constraint $k^{ic}_1$ is positive.
One could very well imagine that a different constraint used to determine the height profile
$h(\vec{r}_{\parallel})$ might lead to a bending rigidity that is positive \cite{Tarazona}.
For the integral constraint the two contributions to $k$ from $k_0$ and $k^{ic}_1$ nearly
cancel leading to a value for $k^{ic}$ which is barely negative. Unfortunately,
this makes the value of $k^{ic}$ sensitively dependent on the precise model used
to determine $\rho_1(z)$.

An important point concerns the {\em scaling behavior} of the bending rigidity. 
The expressions in Eq.(\ref{eq:k_SQ_SR}) indicate that the bending rigidity vanishes
near the critical point with the same exponent as the surface tension, i.e.
\begin{equation}
\label{eq:scaling}
k \propto \frac{B \, \sigma}{m} \, \propto \, \sigma \, d^2 \,.
\end{equation}
Note that for the depletion potential, the ratio $B/m$ only depends on the
size ratio parameter $\varepsilon$ ($B/m \approx$ 0.342 $d^2$ for
$\varepsilon\!=\!$ 1.8) but is independent of $\eta_p$ (or $\Delta \eta$),
see the Appendix.
Both contributions to the bending rigidity, $k_0$ and $k_1$, show the above
scaling behavior and both should therefore be taken into account.

The scaling result in Eq.(\ref{eq:scaling}) should be contrasted with the usual
assumption that $k \!\propto\! \sigma \, \xi^2$, i.e. the bending rigidity
approaches a finite, non-zero limit at the critical point. This scaling
behavior is, for instance, obtained for the bending rigidity $k_{\rm eq}$
determined from analysing the surface tension of a spherically or cylindrically
shaped liquid droplet in metastable equilibrium with a bulk vapor \cite{Blokhuis93, Gompper}.
In the gradient expansion, one has \cite{Blokhuis93}:
\begin{equation}
k_{\rm eq} = - \frac{1}{9} (\pi^2 - 3) \, m \, (\Delta \rho)^2 \, \xi \,.
\end{equation}
It is perhaps important to discuss more broadly this result in the
context of previous work on the virial approach \cite{Blokhuis92}
to the bending rigidity (and other curvature parameters). The virial
expression for the bending rigidity is generally valid, but it is
important to realise that it features the way in which the pair
density depends on curvature \cite{Blokhuis92}. When a mean-field,
squared-gradient approximation is subsequently made \cite{Blokhuis93},
this translates into the expression for the bending rigidity to
depend on the way in which the {\em density} depends on curvature,
i.e. it features the profile $\rho_1(z)$. Therefore,
even though the expressions for the bending rigidity in the
equilibrium approach and the fluctuating interface approach are
the same, they might lead to different values (and scaling behavior)
of the bending rigidity due to the fact that the density profiles
$\rho_1(z)$ are different in these two cases.

It is interesting to also compare with the approach by Mecke and Dietrich \cite{Mecke}.
Even though the goal in ref.~\onlinecite{Mecke} is to consider long-ranged forces, one may use
the expression in Eq.(\ref{eq:rho1_MD}) for $\rho^{\rm MD}_1(z)$ inserted into
Eq.(\ref{eq:Delta_Omega}) to determine the leading correction to $\sigma(q)$ also
for {\em short-ranged} forces. The gradient expansion then gives:
\begin{equation}
\label{eq:k_MD}
k_{\rm MD} = - \left[ \, \frac{C_{\rm H}}{6} - \frac{C_{\rm H}^2}{24 \, \pi^2} \,
( \pi^2 + 16 ) \, \right] m \, (\Delta \rho)^2 \, \xi \,.
\end{equation}
The prefactor is negative (as long as $0 \!<\! C_{\rm H} \!<\! 1.526$) in line
with the results obtained here. Again, the scaling behavior -- equal to that of
$k_{\rm eq}$ -- is essentially different than our prediction in Eq.(\ref{eq:scaling}).

Finally, we like to mention an expression for the bending rigidity that is derived
from the generally valid virial expression \cite{Blokhuis92}, in which the assumption
is made that the width of the interfacial profile is much smaller than the molecular
diameter $d \!\gg\! \xi$ \cite{Blokhuis92, Napiorkowski}. The implication is that
$\rho_0(z)\!=\!\rho_{\rm step}(z)$, the sharp-profile approximation, and $\rho_1(z)\!=\!0$.
The sharp-profile expressions for the surface tension (also known as the Fowler
formula \cite{Fowler}) and bending rigidity are \cite{Fowler, Blokhuis92, Blokhuis99}:
\begin{eqnarray}
\sigma_{\rm sp} &=& \frac{(\Delta \rho)^2}{32} \, \int \!\! d\vec{r}_{12} \; r^2 \, U^{\prime}(r) \, g_{\ell}(r) \,, \\
k_{\rm sp}      &=& - \frac{(\Delta \rho)^2}{768} \, \int \!\! d\vec{r}_{12} \; r^4 \, U^{\prime}(r) \, g_{\ell}(r) \,, \nonumber
\end{eqnarray}
where $U(r)$ is the {\em full} interaction potential. Since $\rho_1(z)\!=\!0$,
the expression for $k_{\rm sp}$ is independent on the constraint used to
determine $\rho_1(z)$.

\begin{figure}
\centering
\includegraphics[angle=270,width=200pt]{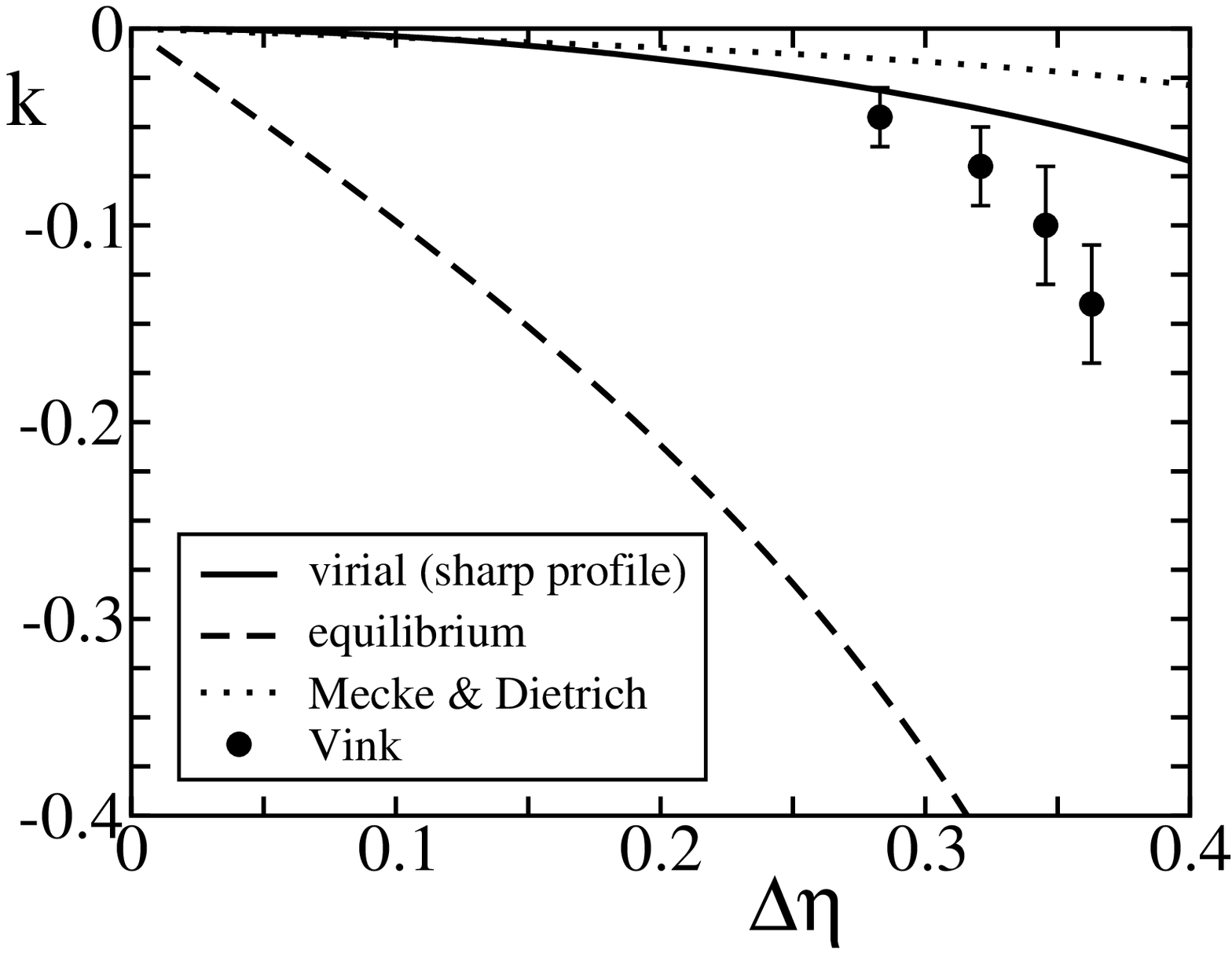}
\caption{Bending rigidity in units of $k_{\rm B} T$
versus the volume fraction difference, $\Delta \eta$.
In this example $\varepsilon \!=\! 1.8$; filled circles
are the MC results by Vink {\em et al.} (Ref.~\onlinecite{Vink}), the drawn
line is the virial expression with the sharp-profile approximation,
the dashed line is the equilibrium result, and the dotted line is
the Mecke and Dietrich result (Ref.~\onlinecite{Mecke})
with $C_{\rm H} \!=\! 1/4$.}
\label{fig:k_SR3}
\end{figure}
In Figure \ref{fig:k_SR3}, the different models for the bending rigidity are
compared to the simulation results of Vink {\em et al.} \cite{Vink}. For
the evaluation of $k_{\rm sp}$ we have taken
$g_{\ell}(r) \!=\! g_{\rm hs}^{\rm PY}(r;\eta_{\ell})$.

\section{DFT: long-ranged interactions}

Here we examine the case that the expansion of $\omega(q,z_{12})$ in $q^2$
{\em cannot} be continued to ${\cal O}(q^4)$. This is the case when the
interaction potential falls of as $1/r^n$ at large distances, with $n\!\leq\!6$.
In particular, we shall assume the asymptotic behavior of $U_{\rm att}(r)$
to be given by 
\begin{equation}
\label{eq:U_as}
U_{\rm att}(r) = - A/r^6 \hspace*{20pt} {\rm when} \hspace*{10pt} r \gg d \,.
\end{equation}
The analysis below only assumes that the asymptotic behavior of $U_{\rm att}(r)$
is given by the above expression. However, when we show explicit results,
we consider the above form for $U_{\rm att}(r)$ extended to the whole range
$1\!<\!r/d\!<\!\infty$ (see also the Appendix).

With the asymptotic behavior of $U_{\rm att}(r)$ given by Eq.(\ref{eq:U_as}),
the expansion of $\omega(q,z_{12})$ in $q^2$ takes on the form:
\begin{eqnarray}
\label{eq:w(q,z12)_LR}
\omega(q,z_{12}) &=& \omega_0(z_{12}) + \omega_2(z_{12}) \, q^2 \\
&& + \frac{\pi}{32} \, A \, q^4 \, \ln(qd) + \omega_4(z_{12}) \, q^4 + \ldots \nonumber
\end{eqnarray}
The coefficient of the $q^4 \ln(q)$-term only depends on the asymptotic
behavior of the interaction potential as defined by the coefficient $A$,
whereas $\omega_4(z_{12})$ depends on the interaction potential's full shape.
With the expansion in Eq.(\ref{eq:w(q,z12)_LR}), $\sigma(q)$ in
Eq.(\ref{eq:sigma_0(q)_k1}) can now be written in the form of
Eq.(\ref{eq:sigma_ECW_LR}) \cite{Mecke}
\begin{eqnarray}
\label{eq:sigma(q)_LR}
\sigma(q) &=& \sigma + k_s q^2 \ln(qd) + k_0 \, q^2 + k_1 \, q^2 + {\cal O}(q^4) \nonumber \\
&\equiv& \sigma + k_s \, q^2 \, \ln(q \ell_k) + {\cal O}(q^4)
\end{eqnarray}
with $k_s \ln(\ell_k/d)\!=\!k_0+k_1$ and  
\begin{eqnarray}
k_s &=& \frac{\pi}{32} \, A \, (\Delta \rho)^2 \,,
\label{eq:k_s} \\
k_0 &=& \int\limits_{-\infty}^{\infty} \!\!\! dz_1 \! \int\limits_{-\infty}^{\infty} \!\!\! dz_{12} \;
\omega_4(z_{12}) \, \rho_0^{\prime}(z_1) \rho_0^{\prime}(z_2) \,,
\label{eq:k0_LR} \\
k_1 &=& - \frac{1}{2} \int\limits_{-\infty}^{\infty} \!\!\! dz_1 \! \int\limits_{-\infty}^{\infty} \!\!\! dz_{12} \;
\omega_2(z_{12}) \, \rho_1(z_1) \rho_0^{\prime}(z_2) \nonumber \\
&& + \frac{\lambda}{2} \int\limits_{-\infty}^{\infty} \!\!\! dz_1 \; \rho_1(z_1) \rho_0^{\prime}(z_1) \,.
\label{eq:k1_LR}
\end{eqnarray}
Next, we proceed to evaluate these expressions in the gradient expansion.

\subsection{Gradient expansion for long-ranged forces}

We first turn to the evaluation of $k_0$ in Eq.(\ref{eq:k0_LR}).
A straightforward gradient expansion of $\rho_0^{\prime}(z_2)$ is now not
possible due to the fact that the integral $\int \! dz_{12} \, \omega_4(z_{12})$
is no longer finite \cite{Mecke}. The assumption of proximity to the critical point,
however, does allow one to consider only the asymptotic form of $\omega_4(z_{12})$
at large distances. Using Eq.(\ref{eq:U_as}), one finds for $\omega_4(z_{12})$
as defined by Eq.(\ref{eq:w(q,z12)_LR})
\begin{equation}
\omega_4(z_{12}) = \frac{\pi A}{32} \, \left[ \gamma_{\rm E} - \frac{3}{4}
+ \frac{1}{2} \ln \!\! \left( \frac{z_{12}^2}{4d^2} \right) \right]
+ {\cal O} \! \left( \! \frac{d^2}{z_{12}^2} \! \right)
\end{equation}
One now proceeds by inserting the above expression for $\omega_4(z_{12})$,
together with the $\tanh$-profile for $\rho_0(z)$ in Eq.(\ref{eq:tanh}),
into the expression for $k_0$ in Eq.(\ref{eq:k0_LR}) and carrying out the
remaining integrations over $z_1$ and $z_{12}$. One finds for $k_0$
\begin{equation}
\label{eq:k0_SQ_LR}
k_0 = \frac{\pi}{32} \, A \, (\Delta \rho)^2 \, \left[ \, \ln(\xi/d) + c_0 + {\cal O}(\frac{d}{\xi}) \, \right] \,,
\end{equation}
where
\begin{eqnarray}
\label{eq:c_0}
&& \!\!\!\! c_0 = \gamma_{\rm E} -\frac{3}{4} - \int\limits_{0}^{\infty} \!\! dt \ln(t^2) 
\left[ \frac{\sinh(t) - t \cosh(t)}{\sinh^3(t)} \right] \nonumber \\
&& \hspace*{5pt} \approx -0.605270 \ldots
\end{eqnarray}
Next, we turn to the evaluation of $k_1$. In the gradient expansion, the expression
for $k_1$ in Eq.(\ref{eq:k1_LR}) reduces to:
\begin{equation}
\label{eq:k1_SQ}
k_1 = - 2 \, m \int\limits_{-\infty}^{\infty} \!\!\! dz \; \rho_1(z) \, \rho_0^{\prime}(z) \,.
\end{equation}
The further evaluation of $k_1$ requires one to solve the Euler-Lagrange equation in
Eq.(\ref{eq:EL_1_SQ}) for $\rho_1(z)$. Again, a gradient expansion of $\rho_0^{\prime}(z_2)$
is not possible due to the fact that now the integral $\int \! dz_{12} \, z_{12}^2 \, \omega_2(z_{12})$
is no longer finite. Using the expression for the interaction potential in Eq.(\ref{eq:U_as}),
one finds for $\omega_2(z_{12})$ when $|z_{12}|\!\gg\!d$
\begin{equation}
\omega_2(z_{12}) = \frac{\pi A}{8 \, z_{12}^2} + {\cal O} \! \left( \frac{d^4}{z_{12}^4} \right) \,.
\end{equation}
The above expression for $\omega_2(z_{12})$ is used to solve Eq.(\ref{eq:EL_1_SQ})
for $\rho_1(z)$ which is then inserted into the expression for $k_1$ in Eq.(\ref{eq:k1_SQ}).
After some algebra, one finally obtains for $k_1$
\begin{eqnarray}
\label{eq:k1_SQ_LR}
k_1^{cc} &=& \frac{\pi}{32} \, A \, (\Delta \rho)^2 \, \left[ \, c^{cc}_1 + {\cal O}(\frac{d}{\xi}) \, \right] \,, \nonumber \\
k_1^{ic} &=& \frac{\pi}{32} \, A \, (\Delta \rho)^2 \, \left[ \, c^{ic}_1 + {\cal O}(\frac{d}{\xi}) \, \right] \,.
\end{eqnarray}
with
\begin{eqnarray}
\label{eq:c_1}
c^{cc}_1 &\approx& -0.559665 \ldots \,, \nonumber \\
c^{ic}_1 &=& \frac{4}{3} \, \int\limits_{0}^{\infty} \!\! dt \; \ln(t) \nonumber \\
&& \hspace*{10pt} \times \left[ \, \frac{\sinh^3(t) + 3 \sinh(t) - 3 t \cosh(t)}{t^2 \sinh^3(t)} \, \right] \nonumber \\
&\approx& 1.461525 \ldots
\end{eqnarray}
\begin{figure}
\centering
\includegraphics[angle=270,width=200pt]{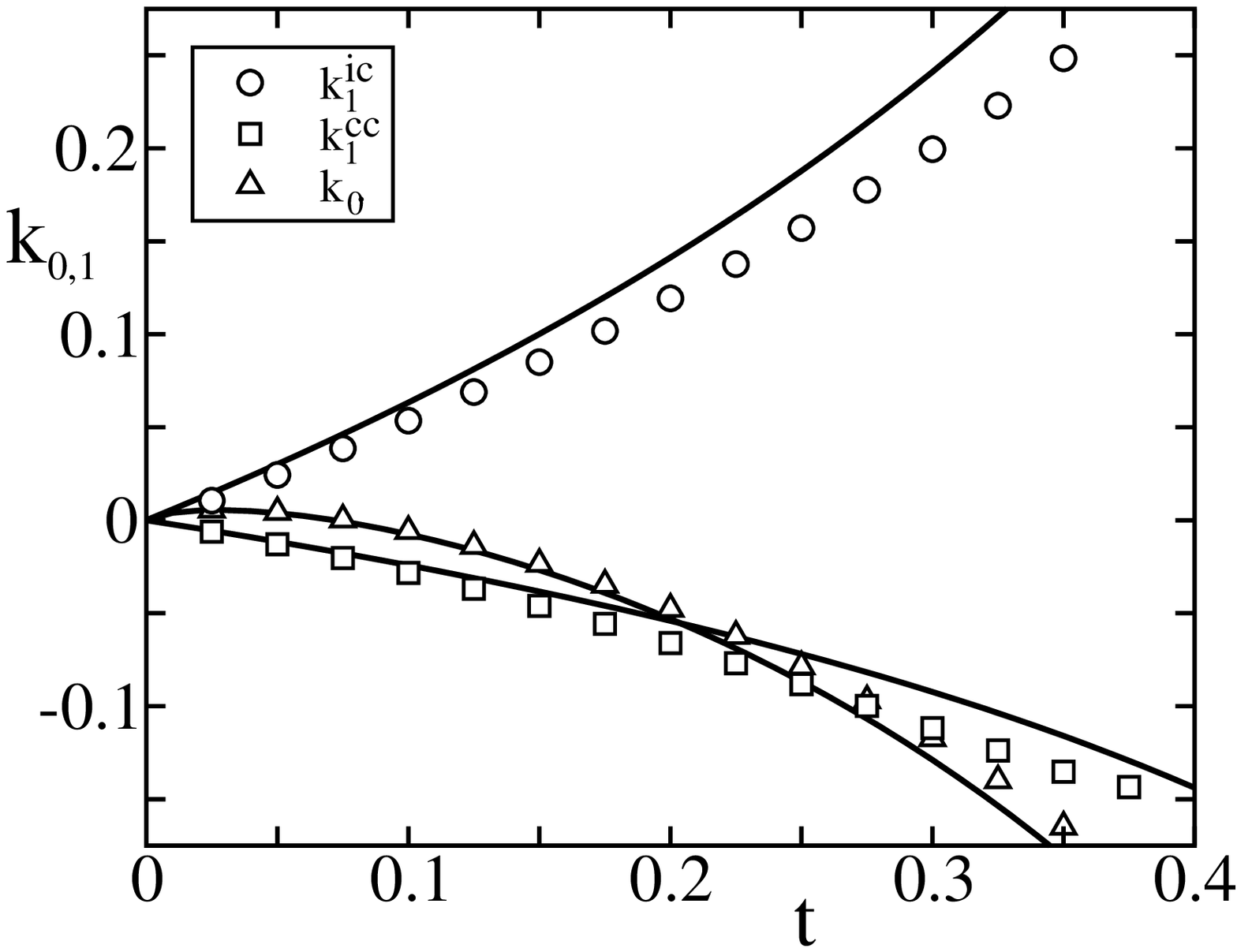}
\caption{Contributions to the bending rigidity $k_0$ and $k_1$ in
units of $k_{\rm B} T$ versus the reduced temperature
distance to the critical point, $t$. Symbols are numerical
results, the drawn lines are the gradient expansion approximation.}
\label{fig:k_GW1}
\end{figure}
\noindent
In Figure \ref{fig:k_GW1}, $k_0$, $k_1^{ic}$ and $k_1^{cc}$ are shown as
a function of the reduced temperature distance to the critical point,
$t\!\equiv\!|1-T/T_c|$. The open symbols are obtained from numerically solving
Eqs.(\ref{eq:EL_0}) and (\ref{eq:EL_1}) to obtain $\rho_0(z)$ and $\rho_1(z)$
and inserting the result into Eqs.(\ref{eq:k0_LR}) and (\ref{eq:k1_LR}).
The drawn lines are the gradient expansion approximation for $k_0$ in
Eq.(\ref{eq:k0_SQ_LR}) and $k_1$ in Eq.(\ref{eq:k1_SQ_LR}).

Adding the expressions for $k_0$ in Eq.(\ref{eq:k0_SQ_LR}) and $k_1$ in
Eq.(\ref{eq:k1_SQ_LR}), one obtains for $\ell_k$
\begin{eqnarray}
\label{eq:k_SQ_LR}
\ell^{cc}_k &=& \exp(c_0 + c^{cc}_1) \, \xi + {\cal O}(d) \nonumber \\
&\approx& 0.311942 \ldots \, \xi + {\cal O}(d) \,, \nonumber \\
\ell^{ic}_k &=& \exp(c_0 + c^{ic}_1) \, \xi + {\cal O}(d) \nonumber \\
&\approx& 2.354329 \ldots \, \xi + {\cal O}(d) \,.
\end{eqnarray}
\begin{figure}
\centering
\includegraphics[angle=270,width=200pt]{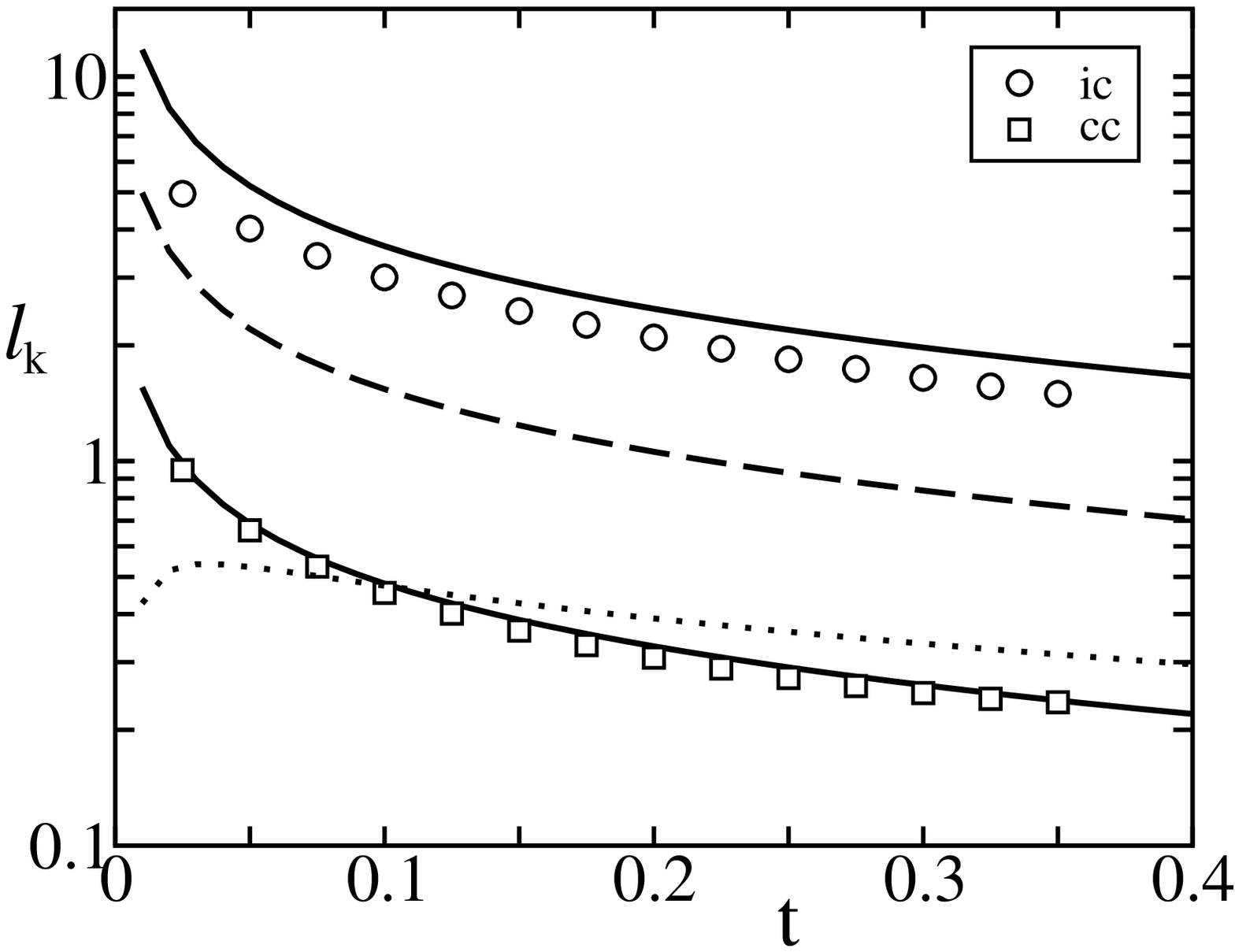}
\caption{The bending length $\ell_k$ (in units of $d$) versus the reduced
temperature distance to the critical point, $t$, using the integral
constraint (circles) and crossing constraint (squares). Open symbols are
numerical results, the drawn lines are the gradient expansion approximation.
The dashed line is the correlation length $\xi$ and the dotted line is
the Mecke and Dietrich result (Ref.~\onlinecite{Mecke}) with $C_{\rm H}\!=\!1/4$.}
\label{fig:k_GW2}
\end{figure}
\noindent
In Figure \ref{fig:k_GW2}, the bending length is shown as a function
of $t$. The open symbols are the numerical results. The drawn lines
are the gradient expansion approximations in Eq.(\ref{eq:k_SQ_LR}).

Other than that the gradient expansion seems not to be as accurate in
reproducing numerical results, the results in Figures \ref{fig:k_GW1}
and \ref{fig:k_GW2} are in line with the earlier results obtained
for short-ranged forces. The leading order correction to the surface
tension is negative when $q \ell_k \!<\! 1$, and the effect is more
pronounced for the crossing constraint than the integral constraint.
Although the goal of Mecke and Dietrich in ref.~\onlinecite{Mecke} is to include
higher order terms, terms beyond $q^2$ in the expansion of $\sigma(q)$,
it is also interesting to compare with the Mecke and Dietrich approach
for the terms obtained to order $q^2 \, \ln(q)$ and $q^2$. Using the
Mecke and Dietrich expression for $\rho_1(z)$ in Eq.(\ref{eq:rho1_MD}),
one has to leading order in the gradient expansion:
\begin{eqnarray}
\sigma_{\rm MD}(q) \!\! &=& \!\! \sigma + k_s \, q^2 \ln(qd) + k_0 \, q^2 + k_{\rm MD} \, q^2 + .. \nonumber \\
&\equiv& \!\! \sigma + k_s \, q^2 \, \ln(q \ell^{\rm MD}_k) + {\cal O}(q^4) \,,
\end{eqnarray}
where $k_s$ and $k_0$ are given by the expressions in Eq.(\ref{eq:k1_LR})
and Eq.(\ref{eq:k0_SQ_LR}), and where $k_{\rm MD}$ is given by the previously
derived expression in Eq.(\ref{eq:k_MD}) for short-ranged forces. In Figure
\ref{fig:k_GW2}, we show, as the dotted line, the result for the bending
length $\ell_k^{\rm MD}$.

\section{Discussion}

In the first part of this article, we have demonstrated that the full spectrum
of surface fluctuations obtained in Monte Carlo simulations \cite{Vink} of the
colloid-polymer interface, can very accurately be described by the following expression:
\begin{equation}
S(q) = \frac{k_{\rm B} T}{\sigma \, q^2} - \frac{k \,\, k_{\rm B} T}{\sigma^2}
+ {\cal N}_L \, S_b(q) \,. 
\end{equation}
The three terms in this expression work in three different $q$-regimes:
\begin{enumerate}
\item{Classical capillary wave regime, $qd \!\ll\! 1$}
\item{Extended capillary wave regime, $qd \!\apprle\! 1$}
\item{Bulk-like fluctuations regime, $qd \!\apprge\! 1$}
\end{enumerate}
Two adjustable parameters are present: ${\cal N}_L$ that weighs
the bulk-like fluctuations compared to the capillary wave
fluctuations and the bending rigidity $k$ as defined by the
leading order correction in an expansion of $\sigma(q)$ in $q^2$,
$\sigma(q) \!=\! \sigma + k \, q^2 + \ldots$. We found that a fit
to the simulation results yields $k \!<\! 0$, i.e. the leading
order curvature correction tends to {\em lower} the surface
tension $\sigma(q)$. This effect is termed {\em capillary enhancement}
\cite{Daillant}; capillary waves are ``more violent'', less restricted
by surface tension, at smaller wavelengths.

One could worry whether a negative bending rigidity is consistent with
having a stable interface. Tarazona {\em et al.} \cite{Tarazona} indicate
that a decrease of $\sigma(q) \!=\! \sigma + k \, q^2 + \ldots$
ultimately leads to a destabilisation of the interface at
{\em large} $q$. It is therefore important to realise that the
extension of the capillary wave model, through the inclusion of
a bending rigidity, is valid only for {\em low} $q$. Higher
order terms in the expansion in $q$ are not systematically
included. In this article we propose to describe $S(q)$
for large $q$ ($qd\!\apprge\!1$) in terms of molecular, bulk-like
fluctuations through $S_b(q)$. It is shown that the full $S(q)$,
which is then a combination of the extended capillary wave model
at low $q$ and bulk-like fluctuations at large $q$, remains
well-behaved ensuring the stability of the interface.
For systems with a low (or even zero) surface tension, the bending
rigidity is the dominant contribution near $q\!=\!0$ and one
necessarily requires a positive value for $k$ \cite{Safran},
but for the simple, (quasi) one-component system considered
here this is not an issue.

A most important and generally underappreciated point that we like
to emphasize is that the location of the interface cannot be defined
{\em unambiguously}. A certain procedure must always be formulated
to determine the height function $h(\vec{r}_{\parallel})$.
We have shown that different choices for the location of the interface,
which are all equally legitimate as long as they lead to a location
of the dividing surface that is `sensibly coincident' with the interfacial
region \cite{Gibbs}, lead to different results for the bending correction
to the capillary wave model. Naturally, all {\em experimentally} measurable
quantities {\em cannot} depend on the chosen location of the interface,
making it necessary to formulate precisely the quantity that is determined
in experiments or simulations. It was shown that for the simulation results,
the value of the bending rigidity in the above expression for $S(q)$,
corresponds to the height function being defined according to the
integral constraint, $k \!=\! k^{ic}$.

For the determination of $k$, it is necessary to take the contribution
from bulk-like fluctuations into account since they also contribute as
a constant, ${\cal N}_L \, S_b(0)$, in the capillary wave regime ($qd\!\ll\!1$).
This observation is consistent with the interpretation of light scattering
results by Daillant and coworkers \cite{Daillant}. To determine $\sigma(q)$,
they subtract from $S(q)$ a contribution proportional to the penetration depth
($\propto {\cal N}_L$) times the liquid compressibility ($\propto S_b(0)$).
Even though the light scattering results by Daillant \cite{Daillant} are
obtained for real fluids, for which the interaction potential is not necessarily
short-ranged, one expects that a description in terms of the above mentioned
three regimes is again useful. A further comparison with the light scattering
results is, however, necessary.

In the second part of this article, a molecular theory to describe
the inclusion of the bending rigidity correction to the capillary
wave model is presented. An essential feature of the theory is the
`Ansatz' made in Eq.(\ref{eq:V_ext}) regarding the thermodynamic
conditions used to vary the interfacial curvature. It improves on
earlier choices made in the sense that the bulk densities are equal
to those at coexistence and the density profile is a continuous function
\cite{Blokhuis92, Parry94, Blokhuis99, FJ}.
The theory predicts that the scaling behavior of the bending rigidity
equals that of the surface tension near the critical point
\begin{equation}
k \, \propto \, \sigma \, d^2 \, \propto \, t^{\mu} \,,
\end{equation}
where $\mu \!\approx\! 1.26$ is the usual surface tension critical exponent
(in mean-field $\mu \!=\! 3/2$) \cite{RW}. This new scaling prediction differs
fundamentally from the scaling of the bending rigidity in the `equilibrium
approach', $k_{\rm eq} \!\propto\! \sigma \, \xi^2$. In this approach the
bending rigidity is determined from considering the equilibrium free energy
of spherically and cylindrically shaped liquid droplets, with their
radii varied by changing the value of the system's chemical potential
\cite{Blokhuis93, Gompper, Giessen}.

The negative sign and scaling behavior of the bending rigidity obtained
from the molecular theory are in accord with Monte Carlo simulations.
However, the magnitude of $k$ from the molecular theory,
$\sqrt{-k/\sigma} \approx$~0.13~$d$, is significantly below the
value obtained in the simulations, $\sqrt{-k/\sigma} \approx$~0.47~$d$
(see also Figure \ref{fig:k_SR2}), but we believe this to be due to
simplifications made in the theory rather than a true discrepancy.

\vskip 20pt
\noindent
{\Large\bf Acknowledgment}
\vskip 10pt
\noindent
I am indebted to Dick Bedeaux for arguing with me on this intriguing
topic since already 20 years. My thoughts have furthermore been shaped
by discussions with giants in this field: John Weeks, Ben Widom, and
Bob Evans. I would like to express my gratitude to Richard Vink for
sharing unpublished simulation results and to Daniel Bonn, Didi Derks
and Joris Kuipers for discussions on the colloid-polymer system.

\appendix
\section{Depletion interaction potential}
\setcounter{equation}{0}

The phase-separated colloid-polymer system is effectively treated as a
one-component system considering the colloids only. The colloid-colloid
interaction is then given by a hard sphere repulsion (diameter $d$) with
an attractive depletion interaction \cite{AOVrij} induced by the presence
of polymers (radius $R_{\rm g}$):
\begin{equation}
U_{\rm dep}(r) = \frac{- k_{\rm B} T \, \eta_p}{2 \, (\varepsilon-1)^3}
\left[ \, 2 \, \varepsilon^3 - 3 \, \varepsilon^2 \left( \frac{r}{d} \right)
+ \left( \frac{r}{d} \right)^{\!3} \, \right]
\end{equation}
with $d\!<\!r\!<\!\varepsilon d$ and the size ratio parameter $\varepsilon$
is defined as:
\begin{equation}
\varepsilon \equiv 1 + \frac{2 R_{\rm g}}{d} \,.
\end{equation}

Using this form for $U_{\rm att}(r)$, the coefficients $a$, $m$, $B$ are
readily calculated to yield
\begin{eqnarray}
a &=& k_{\rm B} T \, d^3 \, \eta_p \, \frac{\pi}{12} \, (2 + 6 \, \varepsilon + 3 \, \varepsilon^2 + \varepsilon^3) \,, \nonumber \\
m &=& k_{\rm B} T \, d^5 \, \eta_p \, \frac{\pi}{240} \, (5 + 15 \, \varepsilon + 10 \, \varepsilon^2 + 6 \, \varepsilon^3 \nonumber \\
&& \hspace*{65pt} + 3 \, \varepsilon^4 + \varepsilon^5) \,, \\
B &=& k_{\rm B} T \, d^7 \, \eta_p \, \frac{\pi}{8400} \, (28 + 84 \, \varepsilon + 63 \, \varepsilon^2  \nonumber \\
&& \hspace*{20pt} + 45 \, \varepsilon^3 + 30 \, \varepsilon^4 + 18 \, \varepsilon^5 + 9 \, \varepsilon^6 + 3 \, \varepsilon^7) \,. \nonumber
\end{eqnarray}
One may also determine the functions $w_0(z_{12})$, $w_2(z_{12})$ and $w_4(z_{12})$.
When $|z_{12}|\!<\!d$ one has:
\begin{eqnarray}
&& w_0(z_{12}) = - k_{\rm B} T \, d^2 \, \eta_p \, \frac{\pi}{5} \, (1 + 3 \, \varepsilon + \varepsilon^2) \,, \nonumber \\
&& w_2(z_{12}) = k_{\rm B} T \, d^4 \, \eta_p \, \frac{\pi}{280} \,
\bigl[ 10 + 30 \, \varepsilon + 18 \, \varepsilon^2 \nonumber \\
&& \hspace*{20pt} + 9 \, \varepsilon^3 + 3 \, \varepsilon^4 - 14 \, \left( 1 + 3 \, \varepsilon + \varepsilon^2 \right)
\, \frac{z_{12}^2}{d^2} \bigr] \,, \nonumber \\
&& w_4(z_{12}) = - k_{\rm B} T \, d^6 \, \eta_p \, \frac{\pi}{20160} \times \nonumber \\
&& \Bigl[ \, 5 ( 7 + 21 \, \varepsilon + 15 \, \varepsilon^2
+ 10 \, \varepsilon^3 + 6 \, \varepsilon^4 + 3 \, \varepsilon^5 + \varepsilon^6 ) \nonumber \\
&& \hspace*{10pt} - 9 \, \left( 10 + 30 \, \varepsilon + 18 \, \varepsilon^2 + 9 \, \varepsilon^3 + 3 \, \varepsilon^4 \right)
\, \frac{z_{12}^2}{d^2} \nonumber \\
&& \hspace*{10pt} + 63 \, \left( 1 + 3 \, \varepsilon + \varepsilon^2 \right) \, \frac{z_{12}^4}{d^4} \Bigr] \,.
\end{eqnarray}
When $d\!<\!|z_{12}|\!<\!\varepsilon d$ one has:
\begin{eqnarray}
&& w_0(z_{12}) = \frac{- k_{\rm B} T \, d^2 \, \eta_p}{(\varepsilon-1)^3} \, \frac{\pi}{5} \nonumber \\
&& \hspace*{1pt} \times \left[ \varepsilon^5 - 5 \, \varepsilon^3 \, \frac{z_{12}^2}{d^2}
+ 5 \, \varepsilon^2 \, \frac{|z_{12}|^3}{d^3} - \frac{|z_{12}|^5}{d^5} \right] \,, \nonumber \\
&& w_2(z_{12}) = \frac{k_{\rm B} T \, d^4 \, \eta_p}{(\varepsilon-1)^3} \, \frac{\pi}{280} \nonumber \\
&& \hspace*{1pt} \times\left[ 3 \, \varepsilon^7 - 14 \, \varepsilon^5 \, \frac{z_{12}^2}{d^2} + 35 \, \varepsilon^3 \, \frac{z_{12}^4}{d^4}
- 28 \, \varepsilon^2 \, \frac{|z_{12}|^5}{d^5} \right. \nonumber \\
&& \hspace*{120pt} \left. + 4 \, \frac{|z_{12}|^7}{d^7} \right] \,, \nonumber \\
&& w_4(z_{12}) = \frac{- k_{\rm B} T \, d^6 \, \eta_p}{(\varepsilon-1)^3} \, \frac{\pi}{20160} \nonumber \\
&& \hspace*{1pt} \times \left[ 5 \, \varepsilon^9 - 27 \, \varepsilon^7 \, \frac{z_{12}^2}{d^2} + 63 \, \varepsilon^5 \, \frac{z_{12}^4}{d^4}
- 105 \, \varepsilon^3 \, \frac{z_{12}^6}{d^6} \right. \nonumber \\
&& \hspace*{60pt} \left. + 72 \, \varepsilon^2 \, \frac{|z_{12}|^7}{d^7} - 8 \, \frac{|z_{12}|^9}{d^9} \right] \,.
\end{eqnarray}

\section{London-dispersion forces}
\setcounter{equation}{0}

The following explicit form for $U_{\rm att}(r)$ is considered:
\begin{equation}
U_{\rm att}(r) = \left\{
\begin{array}{cl}
- A/r^6 & \hspace*{10pt} {\rm when} \hspace*{10pt} r > d \,, \\
0       & \hspace*{10pt} {\rm when} \hspace*{10pt} r < d \,.
\end{array}
\right.
\label{eq:London}
\end{equation}
Using this form for $U_{\rm att}(r)$, the coefficients $a$ and $m$
are readily calculated to yield
\begin{equation}
a = \frac{2 \pi A}{3 \, d^3} \hspace*{20pt} {\rm and} \hspace*{20pt}
m = \frac{\pi A}{3 \, d} \,.
\end{equation}
With this form for the interaction potential one may expand $\omega(q,z_{12})$.
When $|z_{12}|\!>\!d$ one has:
\begin{eqnarray}
&& \omega(q,z_{12}) = - \frac{\pi A}{2 \, z_{12}^4} + \frac{\pi A}{8 \, z_{12}^2} \, q^2
+ \frac{\pi A}{32} \, q^4 \, \ln(qd) \nonumber \\
&& + \frac{\pi A}{32} \, q^4 \left[ \, \gamma_{\rm E} - \frac{3}{4}
+ \frac{1}{2} \ln \!\! \left( \frac{z_{12}^2}{4d^2} \right) \, \right] + \ldots
\end{eqnarray}
When $|z_{12}|\!<\!d$ one has:
\begin{eqnarray}
&& \omega(q,z_{12}) = - \frac{\pi A}{2 \, d^4}
+ \frac{\pi A}{8 \, d^2} \, \left( 2 - \frac{z_{12}^2}{d^2} \right) \, q^2 \nonumber \\
&& + \frac{\pi A}{32} \, q^4 \, \ln(qd) \\
&& + \frac{\pi A}{32} \, q^4 \left[ \gamma_{\rm E} - \frac{3}{2} - \ln(2)
+ \frac{z_{12}^2}{d^2} - \frac{z_{12}^4}{4 d^4} \right] + \ldots \nonumber
\end{eqnarray}

\end{document}